\documentclass[a4paper]{iopart}
\pdfoutput=1
\usepackage{textcomp}
\usepackage{amsmath}
\usepackage{amssymb}
\usepackage{graphicx} 
\usepackage{pgf}
\usepackage[showframe=false]{geometry}
\usepackage{mathdots}
\usepackage[pdftex,bookmarks,colorlinks]{hyperref}
\usepackage{hyperref}
\newcommand{\U}{{\bf U}}
\newcommand{\V}{{\bf V}}
\newcommand{\ket}[1]{\left|#1\right\rangle}
\newcommand{\bra}[1]{\left\langle #1\right|}
\newcommand{\ud}{\mathrm{d}}

\newcommand{\mean}[1]{\left\langle #1\right\rangle}

\newcommand{\nep}{\textrm{e}}

\newcommand{\opgamma}[1]{{\hat{\gamma}^{\phantom \dagger}}_{#1}}
\newcommand{\opgammadag}[1]{{\hat{\gamma}^{\dagger}}_{#1}}
\newcommand{\mopgamma}[1]{{\check{\gamma}^{\phantom \dagger}}_{#1}}
\newcommand{\mopgammadag}[1]{{\check{\gamma}^{\dagger}}_{#1}}
\newcommand{\Real}{\operatorname{\Re{\rm e}}}
\newcommand{\Aimag}{\operatorname{\Im{\rm m}}}
\newcommand{\opc}[1]{{\hat{c}^{\phantom \dagger}}_{#1}}
\newcommand{\opcdag}[1]{{\hat{c}^{\dagger}}_{#1}}
\newcommand{\opd}[1]{{\hat{d}^{\phantom \dagger}}_{#1}}
\newcommand{\opddag}[1]{{\hat{d}^{\dagger}}_{#1}}
\newcommand{\mopc}[1]{{\check{c}^{\phantom \dagger}}_{#1}}

\newcommand{\opbfgamma}[1]{{\widehat{\boldsymbol{\gamma}}^{\phantom \dagger}}_{#1}}
\newcommand{\opbfgammadag}[1]{{\widehat{\boldsymbol{\gamma}}^{\dagger}}_{#1}}

\newcommand{\opbfc}[1]{{\widehat{\boldsymbol{c}}^{\phantom \dagger}}_{#1}}
\newcommand{\opbfcdag}[1]{{\widehat{\boldsymbol{c}}^{\dagger}}_{#1}}

\newcommand{\mopd}[1]{{\check{d}^{\phantom \dagger}}_{#1}}

\begin{document}

\title{Entanglement entropy in a periodically driven Ising chain}

\author{Angelo Russomanno$^{1,2,4,5}$, Giuseppe E. Santoro$^{2,3,5}$, Rosario Fazio$^{2,6}$}
\address{$^1$ Scuola Normale Superiore, Piazza dei Cavalieri 7, 56127 Pisa, Italy}
\address{$^2$ ICTP, Strada Costiera 11, 34151 Trieste, Italy}
\address{$^3$ CNR-IOM Democritos National Simulation Center, Via Bonomea 265, 34136 Trieste, Italy}
\address{$^4$ Department of Physics, Bar-Ilan University (RA), Ramat Gan 52900, Israel}
\address{$^5$ SISSA, Via Bonomea 265, 34136 Trieste, Italy}
\address{$^6$ NEST, Scuola Normale Superiore and Istituto Nanoscienze-CNR, 56126 Pisa, Italy}
\eads{\mailto{arussoma@ictp.it}, \mailto{santoro@sissa.it}, \mailto{fazio@ictp.it} }
%
\begin{abstract}
In this work we study the entanglement entropy of a uniform quantum Ising chain in transverse field undergoing a periodic driving of period 
$\tau$. By means of Floquet theory we show that, for any subchain, the entanglement entropy tends asymptotically to a value $\tau$-periodic 
in time. We provide a semi-analytical formula for the leading term of this asymptotic regime: It is constant in time and obeys a volume law. 
The entropy in the asymptotic regime is always smaller than the thermal one: because of integrability the system locally relaxes to a 
Generalized Gibbs Ensemble (GGE) density matrix. {The leading term of the asymptotic entanglement entropy is completely determined by this GGE density matrix}. Remarkably, the asymptotic entropy shows marked features in correspondence to some non-equilibrium quantum phase transitions undergone by a Floquet state analog of the ground state.
\end{abstract}
%
%
\section{Introduction}
Understanding entanglement  in quantum many-body  systems  is a powerful way to unveil their properties (see the 
reviews~\cite{amico08,Eisert_RMP10}). Among all possible ways to quantify non-local correlations in  a many-particle system 
a key role is played by the entanglement entropy~\cite{Nielsen_Chuang:book}. For a system characterized by a (pure state) 
density matrix $\rho=\ket{\psi}\bra{\psi}$ the entanglement entropy is defined -- given a partition into two  
subsystems $A$ and $B$, and a corresponding reduced density matrix $\rho_A=\Tr_B \rho$ --  
as the von Neumann entropy of $\rho_A$, $S_A=-\Tr_A\rho_A\log\rho_A$.

The scaling of $S_A$ with the size of the subsystem $A$ carries distinct information on the state of the system~\cite{Eisert_RMP10,Latorre_PRL03,
Latorre_QIC04,Calabrese_JSTAT04,Sred_PRL71,Jin_JStP04,Jin_JPA05,Peschel_JSTAT04,Durello_PRl94,Calzamaglia_PRl94}. The ground 
state of a short-range Hamiltonian obeys an area law (scaling like the measure of the border of subsystem $A$), with important corrections arising when the 
system approaches a critical point. Excited states, on the other side -- in a similar fashion as thermal states -- follow typically  a volume law, where the scaling is with the measure of subsystem $A$ (a notable exception are many-body localized states~\cite{Richard_PRL12,Chittarospo_JSTAT13}).

Understanding the behaviour of entanglement entropy is also important in the description of the non-equilibrium dynamics 
of closed quantum many-body systems~\cite{Polkovnikov_RMP11}. Numerous examples support this statement.  Entanglement entropy was 
analysed in the adiabatic dynamics of a many-body system passing through a critical point~\cite{Cherng_PRA06,Cincio_PRA07,Caneva_PRB08}.
{On the opposite case of a rapid change of the coupling parameters -- a quantum quench -- Calabrese and Cardy~\cite{Calabrese_JSTAT05} were the
first to study the time-dependence of the entropy in a conformal field theory, specializing then to the case of a one-dimensional integrable
spin chain. 
They found that the entropy increases linearly before saturating to an asymptotic value (that scales with the volume of the subsystem)
(see also Ref.~\cite{Fagotti_PRL08}). }
In the presence of disorder, for a many-body localized state, the entropy increases logarithmically~\cite{Moore_PRL12,Dechiara_JSTAT06,Znidaric_PRB08,Vosk_PRL15,Rajeev_NJP} -- 
a behaviour intimately connected to the dephasing mechanism characteristic of this phase. {For a disordered ergodic system -- on the opposite -- the disorder-averaged entanglement entropy increases linearly in time before saturation, and the asymptotic disorder-distribution
of the entanglement entropy coincides with the thermal one~\cite{Rajeev_NJP}.}
In addition to the interest in its own, the study of the entanglement is also of utmost importance in connection with the possibility to 
numerically study correlated systems by means of classical simulations, as e.g. the density matrix renormalization group~\cite{Schollwoch_RMP05,Latorre_QIC04}. 
Indeed the power of numerical simulations is limited by the amount of entanglement present in the quantum state.

 
We will analyse the entanglement entropy in another class of non-equilibrium dynamics in which the quantum system is subject to a 
periodic driving.
In recent years there has been an uprising interest for periodically driven closed quantum systems, with special emphasis on the
existence of an asymptotic time-periodic steady state and the conditions under which it shows thermal properties~\cite{Lignier_PRL07,Zenesini_PRL09,Gemelke_PRL05,
Eckardt_PRL051,Kollath_PRL06,Ponte_PRL15,Ponte_AP15,Kim_PRE14,Polkovnikov_NatPhys11,Emanuele_2014:preprint,
Dalessio_AP13,Rigol_PRX14,Russomanno_PRL12,Russomanno_JSTAT13,Russomanno_EPL15,Lazarides_PRL14,Lazarides_PRE14,
Batisdas_PRA12,Knappa_preprint15,Rosch_PRA15,mori_prl16,mori_ap16}. 
{The asymptotic state of open driven quantum systems has also been studied in Refs.~\cite{prosen_prl11,Grifoni_PR98,Russomanno_PRB11,Refael_PRX15,Rosch_PRB14,diehl_arx,diehl_arx1,mori_arx}.}

From now on, we will focus our attention on closed driven quantum systems.
By means of Floquet theory~\cite{Hanggi_book,Shirley_PR65,Hausinger_PRA10}, formal similarities between this case and that of 
a quantum quench have been discovered; at the same time, many results first obtained in the case of a quantum quench have been 
generalized to the periodic driving case. One example is the relaxation to the Floquet diagonal ensemble of periodically driven 
systems~\cite{Russomanno_PRL12,Russomanno_JSTAT13,Russomanno_EPL15,Lazarides_PRL14,Lazarides_PRE14,Russomanno:phdthesis}. 

The behaviour of the entanglement entropy has been analysed as well in periodically driven systems. The relation between 
the way entanglement entropy grows in time and the system being many-body localized or ergodic is the same in periodically driven systems and 
quantum quenches~\cite{Ponte_PRL15}. Transitions between regimes with anomalous scaling exponents
in the finite-time entanglement entropy of an integrable fermionic system was discussed in Ref.~\cite{Sen_ar15:preprint}.
{In Ref.~\cite{bingo_ingo_08} the authors study the entanglement entropy of a free electron chain undergoing
a periodic quench, discussing relations to equilibrium lattice models. 
They find the entanglement entropy saturating to an asymptotic value proportional to the volume of the subsystem, 
displaying in general slow oscillations in time. In Ref.~\cite{lukin_08} the authors describe the periodic driving of the Heisenberg model.
They see that the entanglement entropy saturates after a transient to an asymptotic value obeying a volume law. Remarkably, this asymptotic
value is the same as the $T=\infty$-completely mixed density matrix: the periodic driving makes this system thermalize
(due to the absence of energy conservation a periodic driving can induce
only $T=\infty$-thermalization -- see~\cite{Rigol_PRX14,Lazarides_PRE14,Ponte_AP15,Rosch_PRA15,Russomanno_EPL15})). }

In this paper we consider the entanglement properties of an integrable quantum spin chain.  We focus on the entanglement entropy $S_l(t)$ 
of a subchain of length $l$ for a quantum Ising chain in transverse field undergoing a time-periodic driving of period 
$\tau$~\cite{Russomanno_PRL12,Russomanno_JSTAT13}. 
The entanglement entropy increases linearly in time, until in the long-time limit an asymptotic $\tau$-periodic value $S_l^{(\infty)}(t)$
is reached~\footnote{Similar 
phenomenology can be also seen in Fig.1 of Ref.~\cite{Sen_ar15:preprint}. Here the authors focus on phenomena at finite time and do 
not make any comment on relaxation to an asymptotic condition.}. The asymptotic value, at leading order, is linear in $l$: in the asymptotic regime the entropy follows a volume law, {as occurs in the driven integrable system of~\cite{bingo_ingo_08}.
{We go beyond the results of Ref.~\cite{bingo_ingo_08} because -- whatever is the form of the periodic drive -- we are able to show this convergence analytically by means of Floquet theory: we express the asymptotic value in terms of the Floquet states of the driven system.}
%
%
Moreover, we are able to provide
a semi-analytical formula for the coefficient of the leading-order-in-$l$ term: we find that it is not only $\tau$-periodic 
but even time-independent.} 

The system is integrable and locally relaxes to a non-thermal asymptotic 
condition~\cite{Russomanno_PRL12} described by a special form of Generalized 
Gibbs Ensemble (GGE) density matrix~\cite{Russomanno:phdthesis,Lazarides_PRL14} which we term Floquet GGE {(for more results on the GGE in sudden quenches, see~\cite{Jaynes_PR57,Cazal:PRL06,Eckstein_PRL08,Cazal:PRA09,Manmana_PRL,Rigol_PRL07,Kollar_PRA08, 
Polkovnikov_RMP11}; Refs.~\cite{Calabrese_PRL11,Caneva_JSM11} discuss in detail GGE for sudden quenches in the quantum Ising chain).
We find that the leading-order-in-$l$ term in the asymptotic entanglement entropy is completely determined by the Floquet GGE. 
Therefore, by means of the entanglement entropy, we see that relaxation to the GGE occurs at arbitrarily large length scales.}

{In agreement with the absence of thermalization, the asymptotic entropy is always well below the completely mixed 
$T=\infty$ density matrix value $l\log 2$ and depends on the parameters of the driving:
%
we can see marked peaks in the asymptotic entropy in correspondence to some non-equilibrium quantum phase transition of a 
Floquet-state-analog of the ground state~\cite{Emanuele_arXiv15}.}
%
%

The behaviour we observe is peculiar to driven integrable systems. In the driven non-integrable case, on the contrary, there
are two possibilities. If the dynamics is ergodic, then the entanglement entropy increases linearly in time until the completely mixed state 
value is eventually reached~\cite{lukin_08}; in the many-body localized case, instead, the entanglement entropy shows a logarithmic increase~\cite{Ponte_PRL15}.

The paper is organized as follows. In Section~\ref{Ising:sec} we introduce the dynamics of the uniform quantum Ising chain in a time-periodic
transverse field. We discuss the application of Floquet theory to this problem and introduce the form of the Floquet GGE density matrix describing 
the asymptotic periodic condition of local observables. 
Section~\ref{enta_dyna:sec} is the heart of this paper: we evaluate the entanglement entropy in the periodically driven case showing that
the entanglement entropy for any subchain relaxes to an asymptotically periodic value. We show also that the leading term 
of the asymptotic value obeys a volume law; we find a semi-analytical formula for this term and we see that it is completely determined by the Floquet GGE. 
In Section~\ref{numerically:sec} we show some
numerical results corroborating our analysis and in Section~\ref{conclusion:sec} we draw our conclusions. Most technical details and
analytical derivations can be found in the appendices.

While this manuscript was {about to be submitted for publication}, we became aware of similar results obtained by T. Apollaro, G.~M. Palma and J. Marino~\cite{Jamiro_arxiv}.

\section{Periodically driven quantum Ising chain} \label{Ising:sec}
%
In this section we discuss the dynamics of a uniform quantum Ising chain in transverse field; we provide more details in~\ref{Bogoliubov:sec} and \ref{enta_entra:sec}. 
%
The Hamiltonian of the system is
%
%
\begin{equation}  \label{h11}
  \hat{H}(t) =-\frac{1}{2}\sum_{j=1}^{L}\left(J\hat{\sigma}_j^z\hat{\sigma}_{j+1}^z + h(t) \hat{\sigma}_j^x\right)  \;.
\end{equation}
Here, the $\hat{\sigma}^{x,z}_j$ are spins (Pauli matrices) at site $j$ of a chain of length $L$ with boundary conditions which can be periodic 
(PBC) $\hat{\sigma}^{x,z}_{L+1}=\hat{\sigma}^{x,z}_1$ or open (OBC) $\hat{\sigma}^{x,z}_{L+1}=0$, and $J$ is a longitudinal coupling ($J=1$ in the following). 
The transverse field is taken uniform and time-periodic, $h(t)=h(t+\tau)$.
%
This Hamiltonian can be transformed, through a Jordan-Wigner transformation (see Eq.~\eqref{wj} and Refs.~\cite{Lieb_AP61,Pfeuty_AP70}), 
into a quadratic-fermion form. 
At equilibrium and for a homogeneous transverse field, $h(t)=h_0$, the model has two gapped phases, 
a ferromagnetic ($\left|h_0\right|<1$),  and a paramagnetic ($\left|h_0\right|>1$) one, separated by a quantum phase transition at $h_c=1$.
Assume now that the transverse field oscillates periodically, for $t\ge 0$, around the uniform value $h_0$, with some amplitude $A$.
%
%
If we assume PBC {for the spins}, we can quite simplify the analysis: going to $k$-space 
(we discuss more general cases in~\ref{Bogoliubov:sec})
$\hat{H}(t)$ becomes a sum of two-level systems: 
\begin{equation} \label{Ht:eqn}
  \hat{H}(t) =  \sum_k^{\rm ABC}
                    \left(\begin{array}{cc}
			\opcdag{k} & \opc{-k}
		\end{array}\right)
	  \mathbb{H}_k(t)
	\left(\begin{array}{c}
			\opc{k} \\
			\\
			\opcdag{-k}
		\end{array}\right) 
\hspace{2mm} \mbox{with} \hspace{4mm} 
\mathbb{H}_k(t)\equiv \left(\begin{array}{cc}
			\epsilon_k(t) &-i\Delta_k\\
			i\Delta_k&-\epsilon_k(t)
		\end{array} \right)
\end{equation}
where $\epsilon_k(t) = h(t)-\cos{k}$, $\Delta_k=\sin{k}$, and the sum over $k$ is restricted to positive $k$'s of the form 
$k=(2n+1)\pi/L$ with $n=0,\ldots,L/2-1$, {and $L$ even}, corresponding to anti-periodic boundary conditions 
(ABC) for the fermions \cite{Lieb_AP61}, 
{as appropriate to the subsector with an even number of fermions, where the initial ground state $\ket{\psi_{\rm GS}}$ \textcolor{black}{for a chain of finite length $L$}~\footnote{\textcolor{black}{The ground state of the system always lies in the subsector with an even number of fermions \cite{Kells_PRB14}.
When $h_0<1$, this state is almost degenerate (although lower in energy) with the ground state in the sector with odd number of fermions (since the fermion number parity is conserved, the two sectors are always 
decoupled during the dynamics). 
In the thermodynamic limit the two ground states are degenerate: the two symmetry-breaking ones, the ones with $\mean{\hat{\sigma}_j^z}=0$, are superpositions
of these states.
Nevertheless, we always focus on quantities which -- after the Jordan-Wigner and Fourier transforms -- 
can be written as sums over the values of $k$ given by the considered boundary conditions. 
For these observables, the thermodynamic limit is always
perfectly defined: it is immaterial if we consider the ground state in the even sector, or the ground state in the odd sector. 
For instance, we have checked that taking PBC for the fermions in Fig.~\ref{entroppo:fig} would give rise to results indistinguishable from the ones for ABC, and both are very near to the same thermodynamic limit. 
Moreover, in this work, whenever we perform the space-Fourier transform and have a sum over $k$, we perform it as an integral by means of a cubic interpolation. Therefore, $k$ is always in the continuum and -- in some sense -- we always take our quantities in their well-defined thermodynamic limit.}} 
and the ensuing time-evolved state $\ket{\psi(t)}$ live~\cite{Kells_PRB14}} .
We will briefly refer to such a set of $k$, in the following, as $k\in {\rm ABC}$.
The Hamiltonian can be block-diagonalized in each $k$ sector with instantaneous eigenvalues
%
$\pm E_k(t)=\pm\sqrt{\epsilon_k^2(t)+\Delta_k^2}$.
%
%
We assume that at time $t=0$ the coherent evolution starts with the system in the ground state of the Hamiltonian $\hat{H}(t=0)$. 
This ground state has a BCS-like form
%
$\ket{\psi_{\rm GS}} = \prod_{k>0}^{\rm ABC} \ket{\psi_k^0} = \prod_{k>0}^{\rm ABC} \left(v_k^{0}+u_k^{0}\opcdag{k} \opcdag{-k}\right) \ket{0}$,
%
with $v_k^{0}=\cos(\theta_k/2)$ and $u_k^{0}=i\sin(\theta_k/2)$ expressed in terms of an angle $\theta_k$ defined by $\tan{\theta_k} = (\sin{k})/(1-\cos{k})$.
%

The evolution of the system can be naturally described through a Floquet analysis \cite{Russomanno_PRL12,Russomanno_JSTAT13}. 
Details on how to compute Floquet modes and quasi-energies in this case are given in \cite{Russomanno_PRL12,Russomanno_JSTAT13} 
and the related supplementary material. 
In~\ref{Bogoliubov:sec} we show how to extend this picture to the case of OBC or inhomogeneous couplings; 
in this section we focus on the case of PBC {for the spins} because it is more transparent and instructive.
The state of the system at all times can be written in a BCS form
\begin{equation} \label{state:eqn}
  \ket{\psi(t)}=\prod_{k>0}^{\rm ABC} \ket{\psi_k(t)}=\prod_{k>0}^{\rm ABC} \left(v_k(t)+u_k(t)\opcdag{k} \opcdag{-k} \right) \ket{0} \;,
\end{equation}
where the functions $u_k(t)$ and $v_k(t)$ obey the Bogoliubov-De Gennes equations 
\begin{equation} \label{deGennes:eqn}
    i\hbar \frac{d}{dt}\left(\begin{array}{cc}
			u_k(t)\\v_{k}(t)
           		\end{array} \right)
		= \mathbb{H}_k(t)
		\left(\begin{array}{cc}
			u_k(t)\\v_{k}(t)
		\end{array} \right) 
	\;,
\end{equation}
with initial values $v_k(0)=v_k^{0}$ and $u_k(0)=u_k^{0}$.

%
%
The previous analysis applies to a general $h(t)$. For a time-periodic $h(t)$, we can find a basis of states which are $\tau$-periodic 
``up to a phase'' in each $k$-subspace. 
These are the Floquet states $\ket{\psi_k^{\pm}(t)}=\nep^{\mp i \mu_k t}\ket{\phi_k^{\pm}(t)}$, where $\ket{\phi_k^{\pm}(t)}$ 
are $\tau-$periodic, and are called Floquet modes, while the $\pm\mu_k$ are real and are called quasi-energies.
The two quasi-energies have opposite signs because $\mathbb{H}_k(t)$ has vanishing trace~\cite{Russomanno_PRL12,Russomanno_JSTAT13}.
The state of the system at any time $t$ can then be expanded as
\begin{equation} \label{expansion:eqn}
  \ket{\psi_k(t)} = r_k^+ \nep^{-i\mu_k t} \ket{\phi_k^+(t)} + r_k^- \nep^{i\mu_k t} \ket{\phi_k^-(t)} \;,
\end{equation}
where $r_k^{\pm} = \left\langle \phi_k^{\pm}(0) \right. \ket{\psi_k(0)}$ are the overlap factors between the 
initial state $\ket{\psi_k(0)}$ and the Floquet modes $\ket{\phi_k^{\pm}(t)}$. 

{In~\cite{Russomanno_PRL12} we find that local observables $\hat{b}$ in this system
relax to an asymptotic $\tau$-periodic condition. In~\cite{Russomanno:phdthesis}, we show that the asymptotic $\tau$-periodic condition 
is described by a $\tau$-periodic GGE density matrix $\hat{\rho}_{\rm FGGE}(t)$, which we define Floquet Generalized Gibbs Ensemble
(see also~\cite{Lazarides_PRL14}).
We find
\begin{equation} \label{periodic_GGE:eqn}
  \bra{\psi(t)}\hat{b}\ket{\psi(t)}\stackrel{\scriptscriptstyle t\to \infty} {\longrightarrow}\Tr\left(\hat{b}\,\hat{\rho}_{\rm FGGE}(t)\right)
\quad{\rm where}\quad
    \hat{\rho}_{\rm FGGE}(t) = \prod_{k>0}^{\rm ABC}\frac{1}{1+\lambda_k^2}\,\nep^{-\lambda_k\,\hat{n}_{P,k}(t)}\;,
\end{equation}
where $\lambda_k\equiv\log\left(|r_k^-|^2/|r_k^+|^2\right)$ (see Eq.~\eqref{expansion:eqn} for the definition of $r_k^\pm$) 
and $\hat{n}_{P,k}(t)$ are fermionic quasi-particle-number-operators $\tau$-periodic in time. 
There are $L$ different $\hat{n}_{P,k}(t)$ operators (as many as the degrees of freedom), which are in 
involution ($\left\{\hat{n}_{P,k}(t),\hat{n}_{P,k'}(t)\right\}=0$ $\forall$ $k,k'$). 
%
The expectation value of 
the operators $\hat{n}_{P,k}(t)$ is conserved in time. In order to see that, we show the precise definition of these operators. They are defined as 
$\hat{n}_{P,k}(t)\equiv\opgammadag{P,k}(t)\opgamma{P,k}(t)$, where the quasi-particle operators $\opgamma{P,k}(t)$ are a linear combination 
of $\opc{k}$ and $\opcdag{-k}$
\begin{equation} \label{gam_modi:eq}
\left(\begin{array}{c}
  \opgamma{P,k}(t)\\\opgammadag{P,-k}(t)
\end{array}\right)=
\left(\begin{array}{cc}
  u_{P,k}(t)&v_{P,k}(t)\\
  -v_{P,k}^*(t)&u_{P,k}^*(t)
\end{array}\right)
\left(\begin{array}{c}
  \opc{k}\\\opcdag{-k}
\end{array}\right)\,.
\end{equation}
The $\tau-$periodic objects $u_{P,k}(t)$ and $v_{P,k}(t)$ are defined in Eq.~\eqref{psi_components:eqn} (they are the components
of the periodic Floquet mode $\ket{\phi_k^+(t)}$ in the basis $\{\opcdag{k}\opcdag{-k}\ket{0},\,\ket{0}\}$ -- see Eq.~\eqref{expansion:eqn}).
With these definitions, it is not difficult to show that
\begin{equation}
  \bra{\psi(t)}\hat{n}_{P,k}(t)\ket{\psi(t)}=\bra{\psi_k(t)}\hat{n}_{P,k}(t)\ket{\psi_k(t)}=\left|r_k^+\right|^2\,,
\end{equation}
where $r_k^+$ is a constant defined in Eq.~\eqref{expansion:eqn}. As we can see, the Floquet Generalized Gibbs Ensemble Eq.~\eqref{periodic_GGE:eqn}
depends on the initial conditions (encoded in $r_k^{\pm}$) and is never thermal: the system is integrable and there are too many conserved quantities (the $\bra{\psi(t)}\hat{n}_{P,k}(t)\ket{\psi(t)}$) to allow thermalization.} 

{We would like to emphasize that the picture above is strictly analogous to the well known situation of the relaxation to the GGE of an integrable
system with $N$ degrees of freedom after a quantum quench~\cite{Jaynes_PR57,Cazal:PRL06,Eckstein_PRL08,Cazal:PRA09,Manmana_PRL,Rigol_PRL07,Kollar_PRA08, 
Polkovnikov_RMP11,Calabrese_PRL11,Caneva_JSM11}. In this case the system has $N$ (local or quasi-particle-number-like) operators $\hat{I}_\mu$, which are in involution and have an expectation
value constant in time~\cite{Polkovnikov_RMP11}. The system locally relaxes to a GGE density matrix $\hat{\rho}_{\rm GGE}$
which is expressed in terms of the operators $\hat{I}_\mu$ and has a structure very similar to Eq.~\eqref{periodic_GGE:eqn} ($\hat{\rho}_{\rm GGE}=\frac{1}{\mathcal{N}}\nep^{-\sum_{\mu=1}^L\lambda_\mu\hat{I}_\mu}$, where $\lambda_\mu$ are some constants fixed by the initial state).}

{In the next section we are going to see how the Floquet theory allows us to show that the stroboscopic entanglement entropy 
of any subchain attains an asymptotic $\tau$-periodic regime. We find a semi-analytical expression for its leading-order value, which depends only on the information
contained in the Floquet Generalized Gibbs Ensemble.}

\section{Dynamics of the entanglement entropy in the periodically driven Ising chain} \label{enta_dyna:sec}
%
The entanglement entropy can be obtained from the
solutions $u_k(t)$ and $v_k(t)$ of the Bogoliubov-de Gennes equations \eqref{deGennes:eqn}. 
In~\ref{enta_entra:sec} we show how to use these solutions to compute a special $2l\times 2l$ time-dependent 
Toeplitz matrix (Eq.~\eqref{corr_mat_tras_t:eqn}). 
In terms of its $2l$ eigenvalues $\pm\nu_m(t)$, $m=1,\ldots l$, we can construct the entanglement entropy of the subchain as 
$S_l(t)=-\sum_{m=1}^l H(\nu_m(t))$, where we have defined the function
%
$H(x) = -\frac{1+x}{2}\log\left(\frac{1+x}{2}\right)-\frac{1-x}{2}\log\left(\frac{1-x}{2}\right)$.
%
This holds true for a generic time-dependent $h(t)$.

In the case of a periodic driving, $h(t+\tau)=h(t)$, we can say something more. 
By applying the Floquet theory and the Riemann-Lebesgue lemma, we are able to show that the entanglement entropy tends 
towards an asymptotic value which is $\tau$-periodic. 
We find that this asymptotically periodic entanglement entropy can be obtained from the eigenvalues 
a special $2l\times 2l$ $\tau$-periodic Toeplitz matrix (Eq.~\eqref{corr_inf_mat:eqn}) which can be evaluated starting 
from the solutions of the Bogoliubov-de Gennes equations expanded in the Floquet basis (see Eqs.~\eqref{expansion:eqn} and~\eqref{psi_components:eqn}) 
 Eq.~\eqref{deGennes:eqn}. 
Defining the eigenvalues of such a Toeplitz matrix as $\pm\nu_m^{(\infty)}(t)$, $m=1,\ldots l$, the asymptotic entanglement entropy 
turns out to be given by
\begin{equation} \label{S_periodic:eqn}
  S_l^{(\infty)}(t)=-\sum_{m=1}^l H\left(\nu_m^{(\infty)}(t)\right) \;.
\end{equation}
All the details of the derivations of this expression are reported in~\ref{relaxation:sec}. 
The discovery of this relaxation phenomenon is one of the main results of the paper.

An important, we believe, observation is that $S_l^{\,(\infty)}(t)$ obeys a volume law.
With an argument analogous to the one used in Ref.~\cite{Calabrese_JSTAT05}, we show in~\ref{Appendix-semi} that
%
\begin{equation} \label{sel_leading:eqn}
  S_l^{\,(\infty)}(t) =  s^{({\infty})}\, l  + \mathcal{O}(\log l)
\quad{\rm with}\quad 
s^{({\infty})} \equiv  -\frac{1}{\pi}\int_{0}^\pi\ud k\, \left[|r_k^-|^2\log|r_k^-|^2+|r_k^+|^2\log|r_k^+|^2\right] \,,
\end{equation}
where $r_k^\pm$ have been defined in Eq.~\eqref{expansion:eqn}. 
Hence, remarkably, the leading term in the asymptotic $\tau$-periodic entanglement entropy is {\em time-independent}
and proportional to $l$ (with at most logarithmic corrections): 
this is reminiscent of the result valid in the case of a quantum quench, found in Ref.~\cite{Calabrese_JSTAT05}. 
{The coefficient $s^{({\infty})}$ is $1/L$ times the von Neumann entropy of the Floquet Generalized 
Gibbs ensemble density matrix $\hat{\rho}_{\rm FGGE}(t)$
(Eq.~\eqref{periodic_GGE:eqn}). So, the entanglement entropy shows that the system relaxes to the Floquet GGE ensemble
at any arbitrarily long spatial scale $l$.}
We can see that $r_k^\pm$ depend strongly on what is the chosen initial state, and so does $s^{({\infty})}$; 
in particular, we never see thermalization, as we are going to show in the next section where we substantiate our analytical 
formulae with numerical results.


\section{Numerical results} \label{numerically:sec}
Numerically solving the Bogoliubov-de Gennes equations ~\eqref{deGennes:eqn} by means of a 4th-order Runge-Kutta algorithm 
(see for instance~\cite{Recipes:book}),
we can explicitly see the relaxation of the entanglement entropy and the validity of the asymptotic value expansion Eq.~\eqref{sel_leading:eqn}.
In all this section we choose a driving of the form
\begin{equation}
  h(t) = h_0 + A \sin\left(\omega_0 t\right) \;,
\end{equation}
where we assume for simplicity $A>0$, $h_0>0$ and $\omega_0>0$. 
In the presentation of the results we will restrict to stroboscopic times of the form $t_n=n\tau$ with $\tau=2\pi/\omega_0$. 
On one side, this choice simplifies the numerical calculations. On the other side, the asymptotically periodic condition for  
$S_l^{\,(\infty)}(t)$ appears as an approach to an asymptotic constant value $S_l^{\,(\infty)}(0)$, which can be seen more easily. 

\begin{figure}
  \begin{center}
    \begin{tabular}{cc}
      \hspace{-1cm}\resizebox{75mm}{!}{\includegraphics{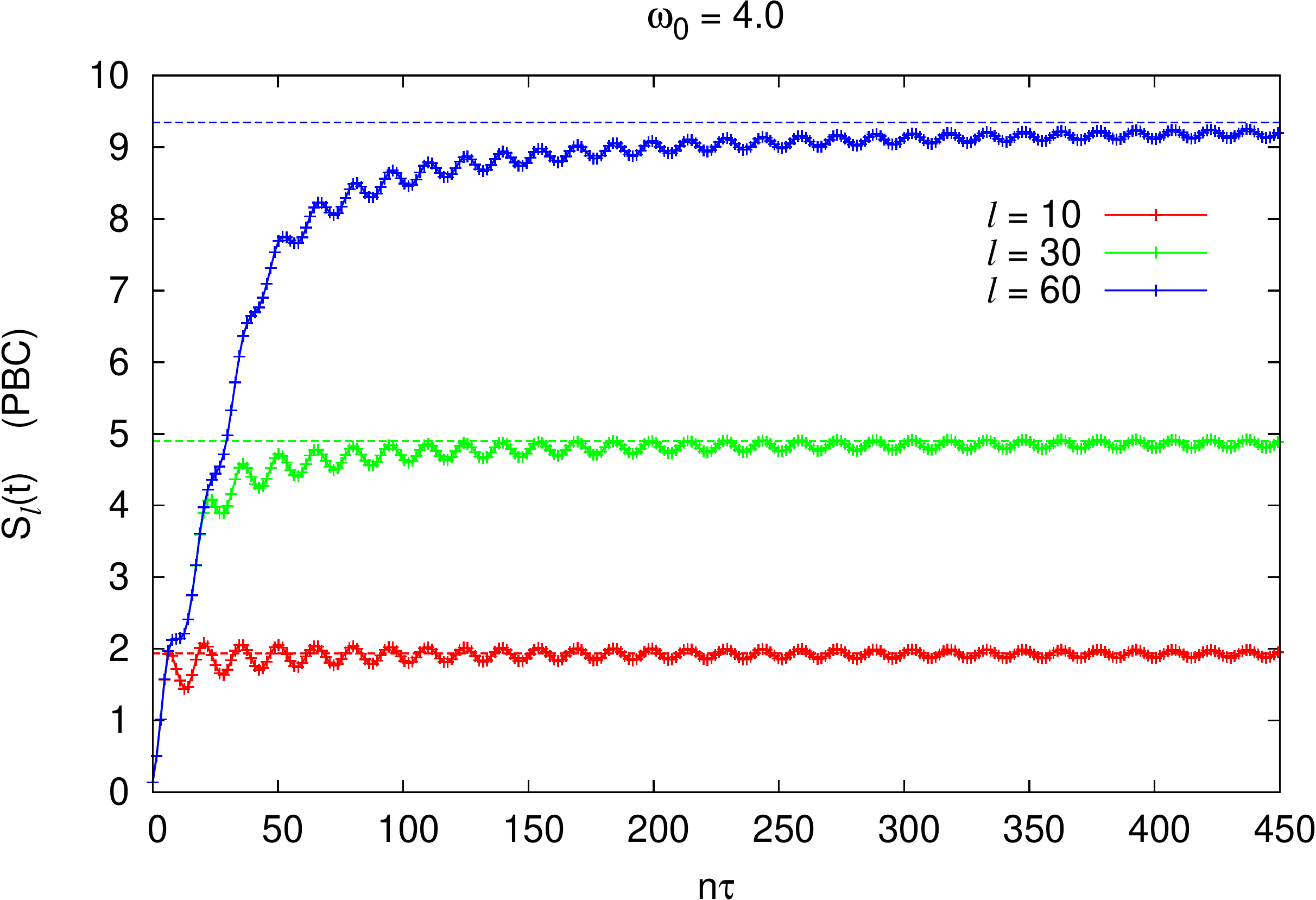}}&
             \resizebox{75mm}{!}{\includegraphics{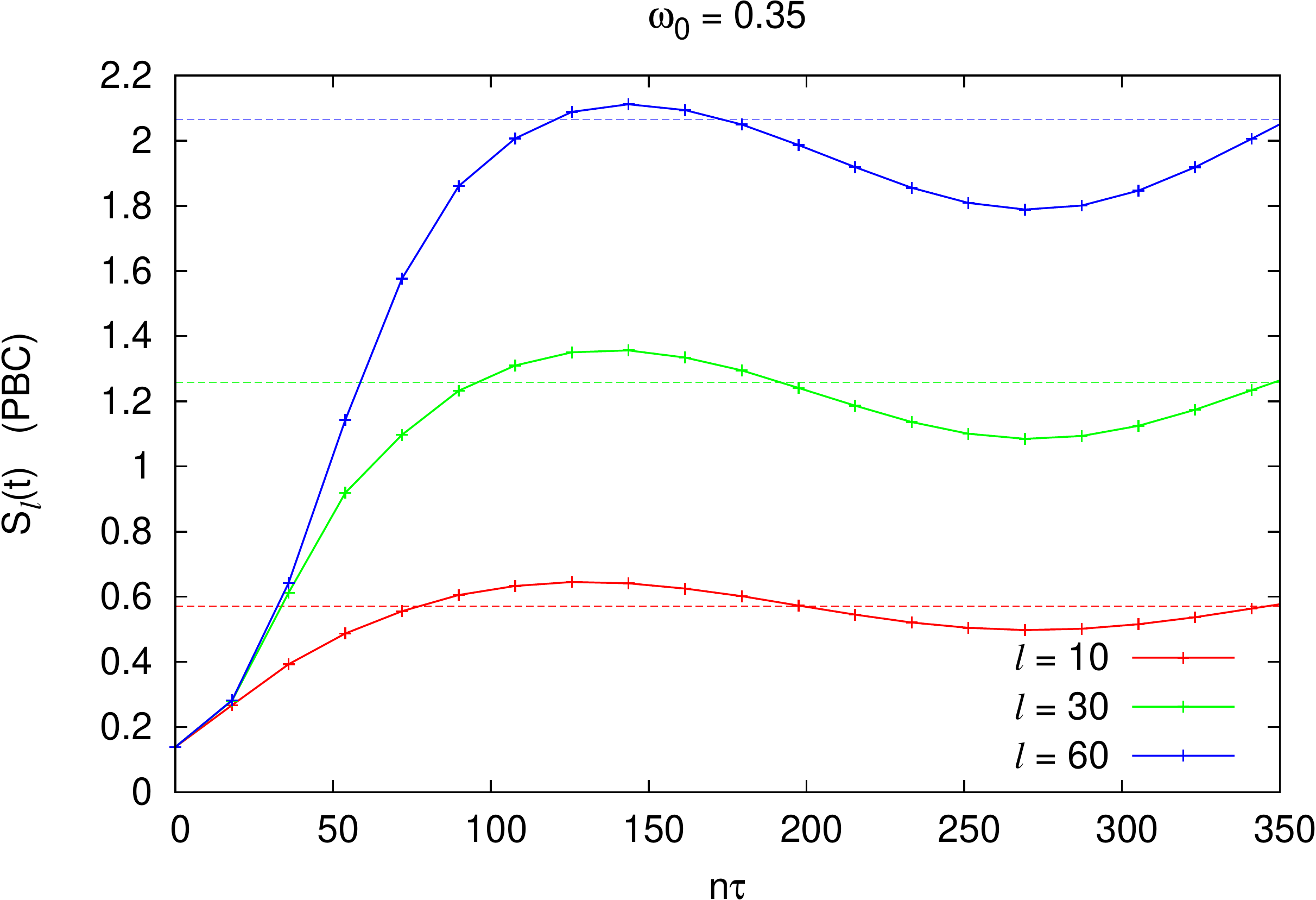}}\\
      \hspace{-1cm}\resizebox{75mm}{!}{\includegraphics{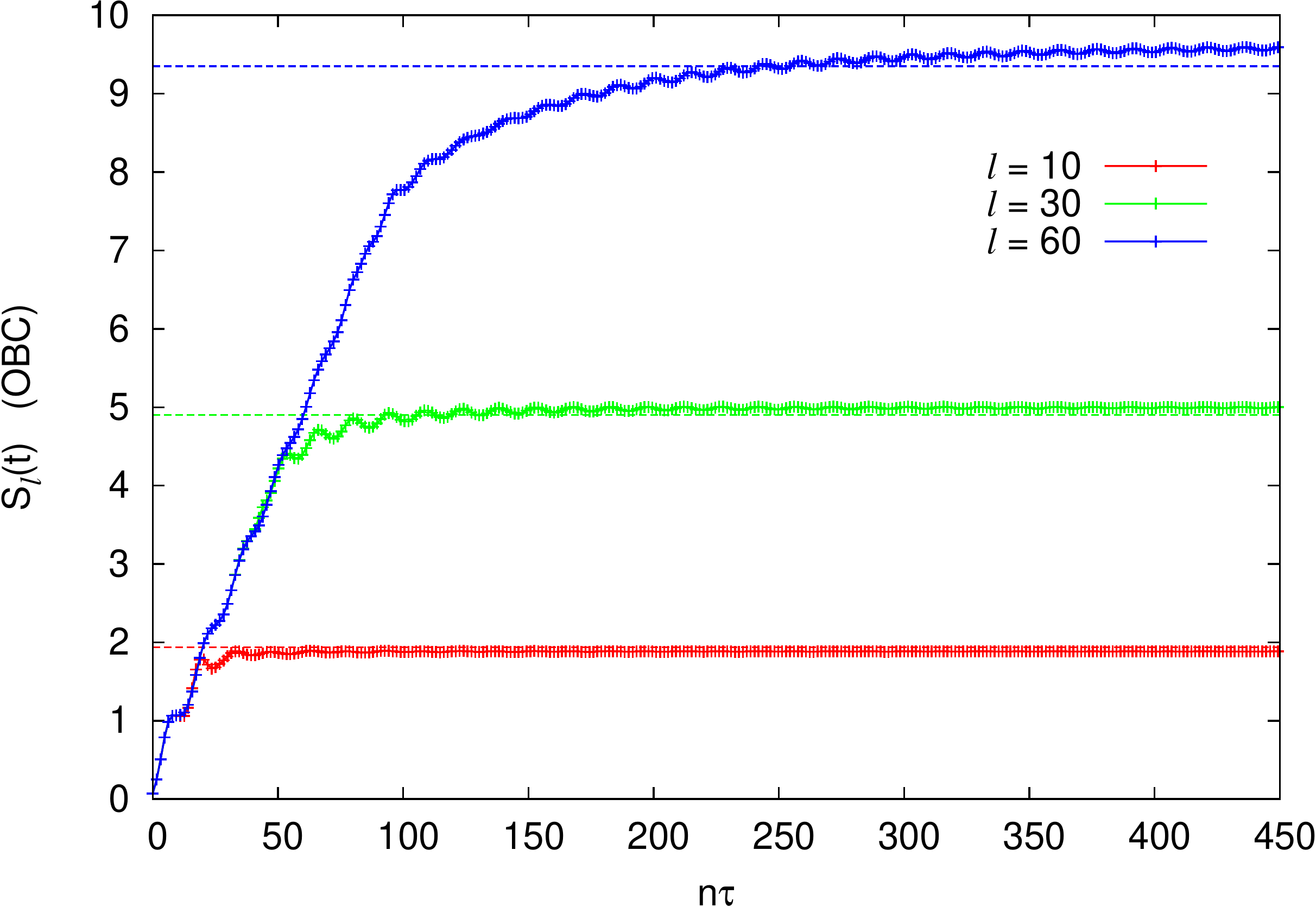}}&
       \resizebox{75mm}{!}{\includegraphics{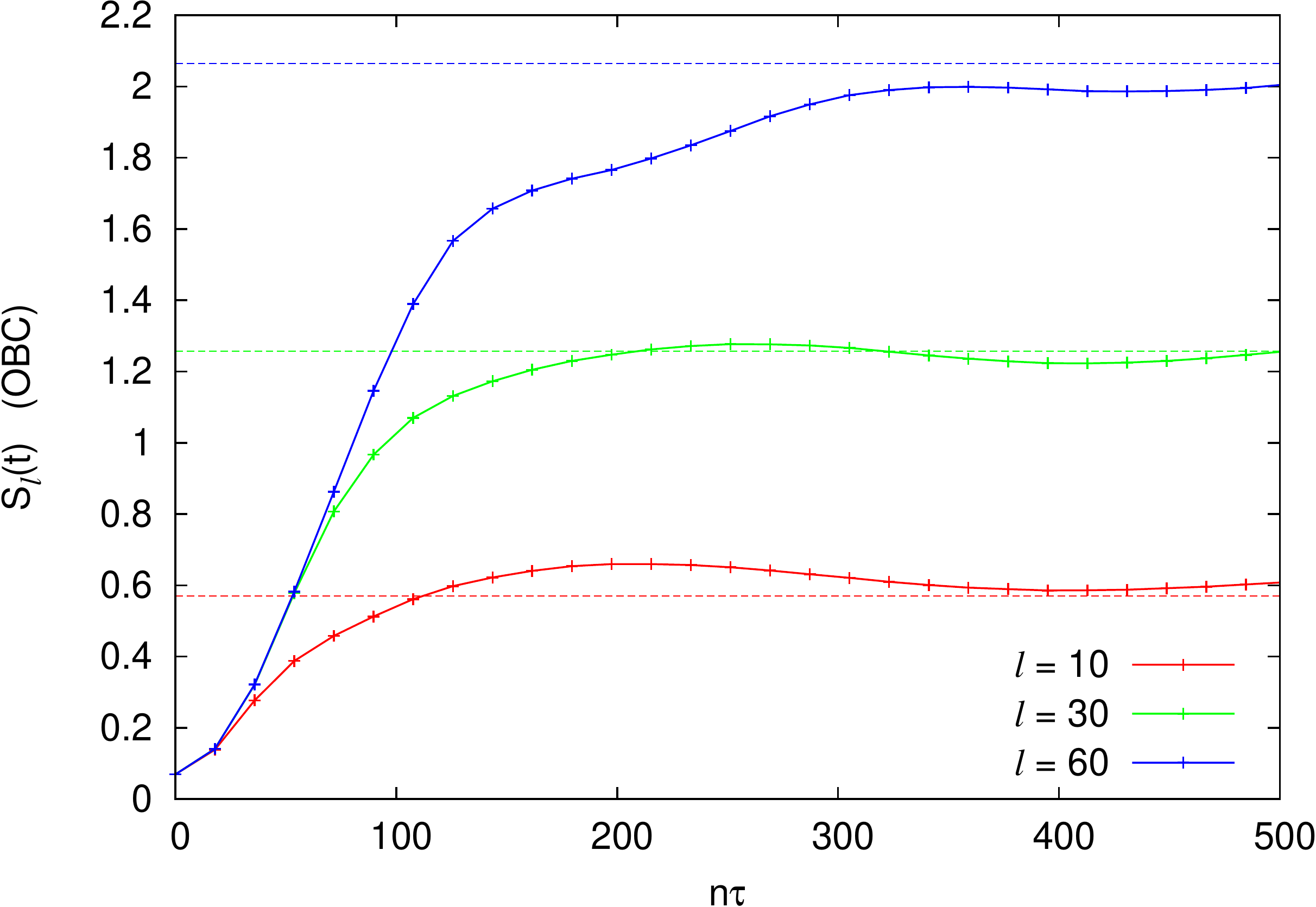}} \\
    \end{tabular}
  \end{center}
\caption{Examples of convergence of the stroboscopic entanglement entropy $S_l(n\tau)$ (solid lines) to the asymptotic value 
$S_l^{\,(\infty)}(0)$ (dashed lines). 
In the top panels we take periodic boundary conditions, in the bottom ones open boundary conditions. 
We cannot follow the convergence for longer times because of finite size revivals (see~\ref{num_imp:sec}).
Numerical parameters: $h_0=2.3$, $L=800$, $A=1.0$; on the left panels it is $\omega_0=4.0$, on the right ones $\omega_0=0.35$.
Details on the numerical methods used for the calculations of these figures are given in~\ref{num_imp:sec}.
}
\label{entroppo:fig}
\end{figure}

We show some examples of stroboscopic relaxation in Fig.~\ref{entroppo:fig}. 
In all the cases we have considered, we see that the (stroboscopic) entanglement entropy reaches, after a transient, its asymptotic value 
$S_l^{\,(\infty)}(0)$. We can see that the transient lasts for a time linearly scaling with $l$. 
It is not difficult to qualitatively understand this fact. In each point of the system the periodic drive 
generates quasi-particle excitations which propagate at finite velocity, bringing with them correlations: 
a longer subchain takes proportionally more time to get fully entangled with the rest of the system. 
This picture strictly resembles the case of quantum quench of~\cite{Calabrese_JSTAT05}. 
It explains why the leading term in the asymptotic entanglement entropy follows a volume law: quasi-particles are generated in
every point of the subchain so, after a while, the whole subchain is correlated with the rest of the system. 
In the top panels of Fig.~\ref{entroppo:fig} we show the situation in the 
case of PBC on the spins (ABC on the fermions), in the bottom ones we consider the case of OBC. 
In the second case the convergence to the asymptotic value is slower (more or less) by a factor 2: 
we are considering a subchain with a loose end, and excitations can propagate only towards one direction. 
Moreover, taking OBC, the oscillations around the asymptotic value seem to be smaller than in the PBC case.
%

\begin{figure}
  \begin{center}
    \begin{tabular}{cc}
      \resizebox{70mm}{!}{\includegraphics{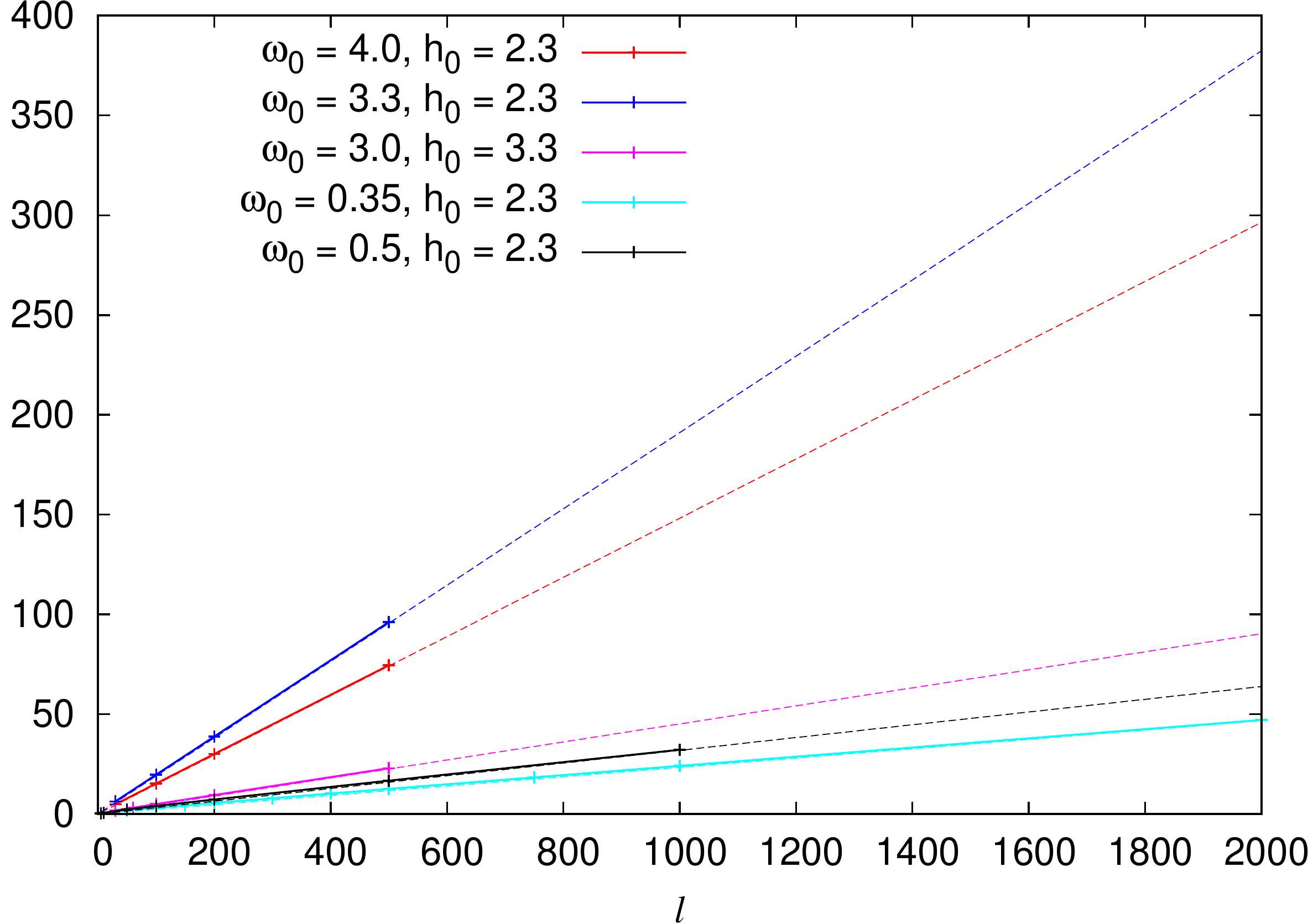}}&
      \resizebox{74mm}{!}{\includegraphics{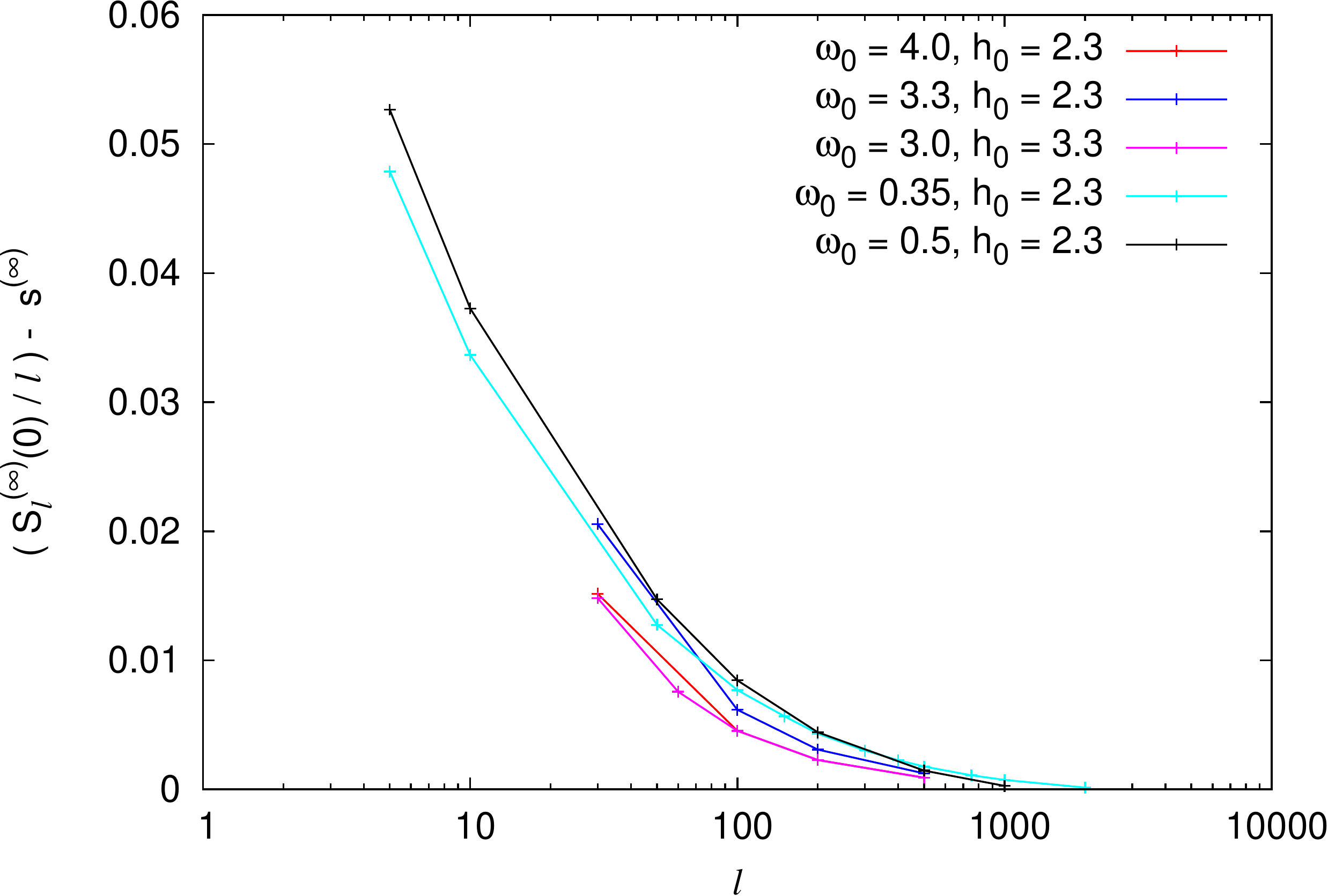}}\\
    \end{tabular}
  \end{center}
\caption{(Left panel) $S_l^{\,(\infty)}(0)$ (solid line) and $l\,s^{\infty}$ vs {the A-subchain length} $l$ (dashed line) 
for some choices of the driving parameters.
(Right panel) $\frac{S_l^{\,(\infty)}(0)}{l}-s^{\infty}$ vs $l$ for some choices of the driving parameters. We see that this difference tends
to 0 for $l\to\infty$: in this limit $l\,s^{\infty}$ is a very good approximation for ${S_l^{\,(\infty)}(0)}$. Numerical parameters: $L=5000$, $A=1.0$. }
\label{linear_l:fig}
\end{figure}

In the left panel of Fig.~\ref{linear_l:fig}, we show the dependence on $l$ of the asymptotic 
entanglement entropy $S_l^{\,(\infty)}(0)$, compared with the semi-analytical result of the leading term $l\,s^{({\infty})}$ 
(Eq.~\ref{sel_leading:eqn}): we see a quite good agreement. 
In the right panel of the same figure we plot the rescaled difference $\frac{S_l^{\,(\infty)}(0)}{l}-s^{({\infty})}$, which clearly 
tends to 0 for $l\to\infty$. Therefore, for $l\to \infty$ we have that $l\,s^{({\infty})}$ is a very good approximation for
$S_l^{\,(\infty)}(0)$. 
%
%

\begin{figure}[b]
  \begin{center}
    \begin{tabular}{cc}
      \resizebox{80mm}{!}{\includegraphics{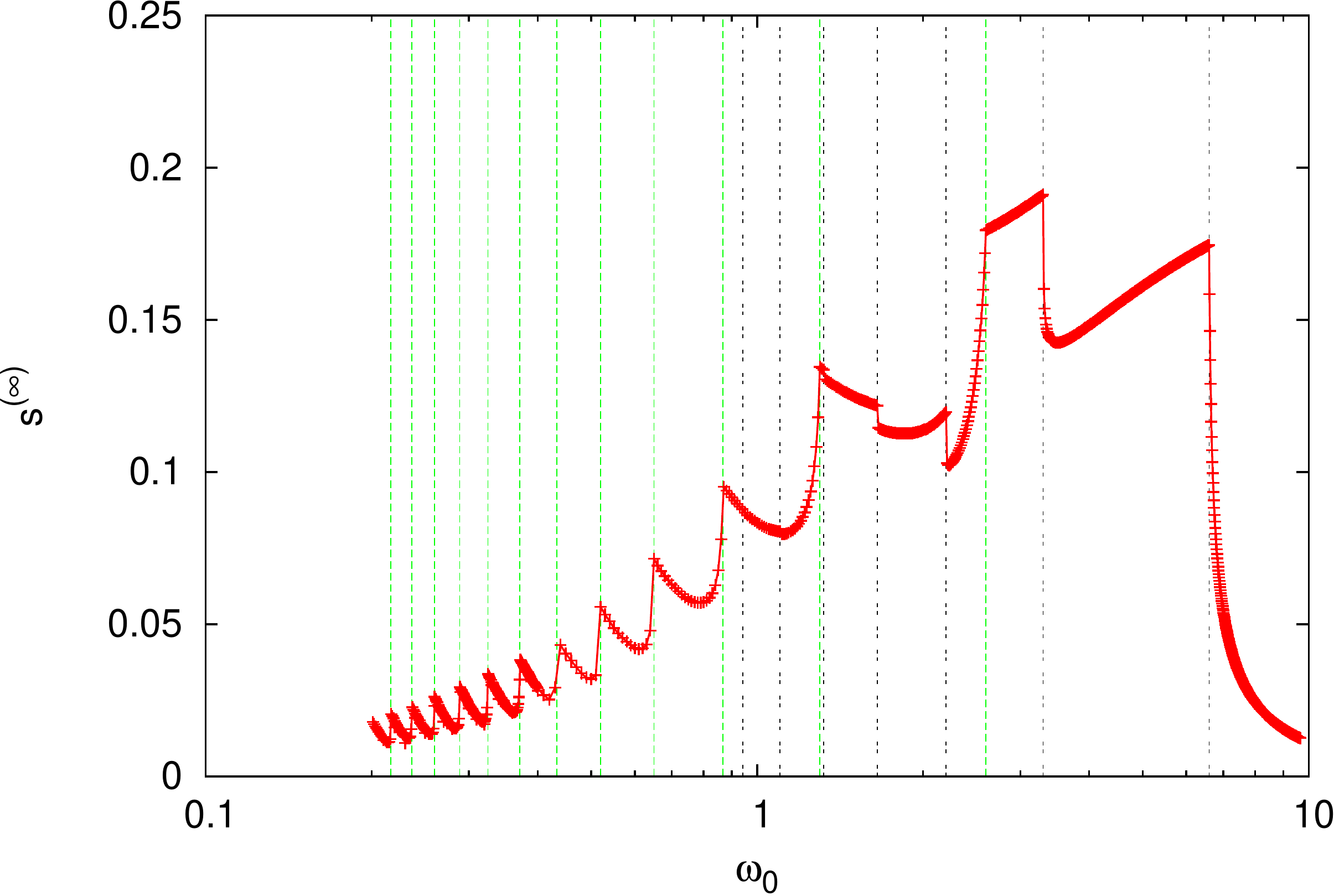}}&
    \end{tabular}
  \end{center}
\caption{The coefficient $s^{(\infty)}$ of the linear term in $l$ of the asymptotic entropy vs $\omega_0$. We see clear peaks at
the frequencies where the $k=0$ mode is resonant ($\omega_0=2|h_0-h_c|/p$ {with $p$ integer}, green vertical lines). 
Instead, at the frequencies where the mode at $k=\pi$ is resonant ($\omega_0=2|h_0+h_c|/q$ {with $q$ integer}, black vertical lines) 
there are much less pronounced features. 
The coefficient is always smaller than its $T=\infty$ thermal value $\log 2$: the system is integrable and never thermalizes whatever the driving parameters.
(Numerical parameters: $A=1.0$, $h_0=2.3$, $L=5000$.)}
\label{sinfomega:fig}
\end{figure}

\begin{figure}
 \begin{center}
  \includegraphics[width=10.0cm]{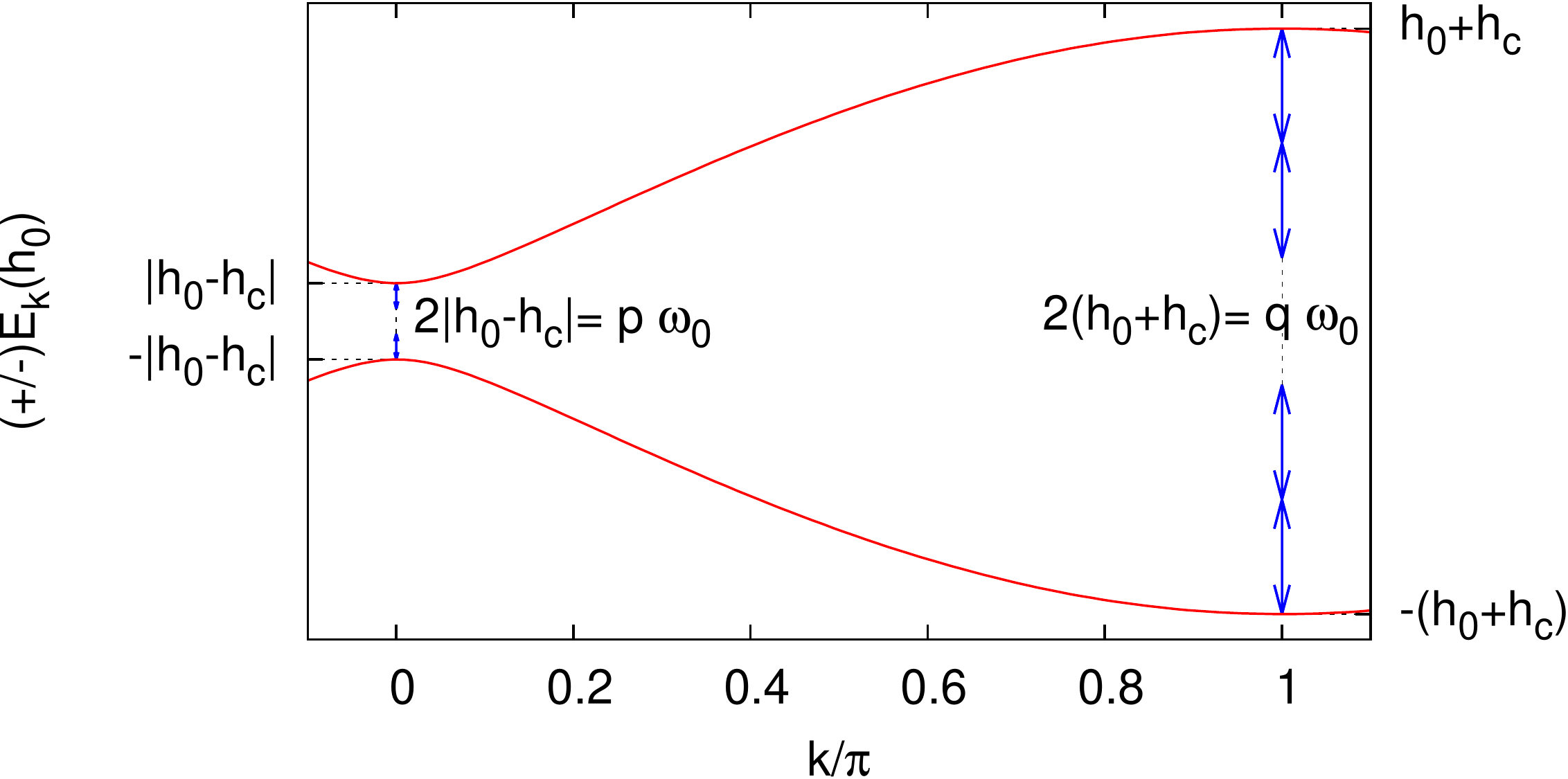}
 \end{center}
  \caption{Schematic explanation of the multi-photon resonances marking the non-equilibrium quantum phase transitions (see main text).}
 \label{a_me_gli_occhi:fig}
\end{figure}

Fig.~\ref{sinfomega:fig} shows the dependence of $s^{({\infty})}$ on $\omega_0$. We see some marked peaks: they appear in
correspondence with a series of Floquet resonances~\cite{russomanno_JSTAT15,Batisdas_PRA12}. 
These resonances are discussed in Ref.~\cite{russomanno_JSTAT15}, 
where it is shown that their positions do not depend on the specific form of the periodic drive $h(t)$ but only on its
time-averaged value $h_0$. 
To better understand these resonances, we have to consider the unperturbed spectrum which is 
made by the two bands (see Fig.~\ref{a_me_gli_occhi:fig}): $\pm E_k\left(h_0\right) = \pm\sqrt{(h_0-\cos{k})^2+\Delta_k^2}$.
In the vanishing amplitude limit ($A\to 0$), if a $k$ mode obeys the relation $n\omega_0=2E_k\left(h_0\right)$, it shows a $n$-photon resonance. 
Moving to the case of a non-vanishing driving amplitude, in general, 
these resonances develop into avoided crossings in the quasi-energy spectrum (see~\cite{Breuer_ZPD89,Emanuele_arXiv15} and the 
Supplemental Material of~\cite{Russomanno_PRL12}). 
The exact degeneracies persist only if they occur at $k=0$ or $k=\pi$.
The $k=0$ resonances occur when $2\left|h_0-h_c\right|=p\,\omega_0$, the $k=\pi$ ones if $2\left|h_0+h_c\right|=q\omega_0$ 
($p,~q\in\mathbb{Z}$ is the order of the resonance -- see Fig.~\ref{a_me_gli_occhi:fig})~\cite{russomanno_JSTAT15}.
We see in Fig.~\ref{sinfomega:fig} very large peaks for the $k=0$ resonances, and much smaller features at the $k=\pi$ 
resonances.~\footnote{As a matter of fact, the transitions between different behaviours of the entanglement entropy at finite time observed in 
Ref.~\cite{Sen_ar15:preprint} occur at resonance points strictly analogous to ours. 
The authors consider a one-dimensional periodically driven fermionic system of the form Eq.~\eqref{Ht:eqn}: they
see transitions in the scaling of the finite-time entanglement entropy at those frequencies where one 
more extremum in the Floquet spectrum $\mu_k$ appears. Such extrema appear right at our resonance points (see SM of~\cite{Russomanno_PRL12}.}
As discussed in Ref.~\cite{Emanuele_arXiv15}, these resonances correspond to second-order quantum
phase transitions of a special Floquet state (the so-called ``Floquet ground state'') 
whose behaviour is analogous to that of the ground state of a physical system. 
Although we are considering the entanglement entropy of a state very different from the Floquet ground state, we remarkably see 
in the behaviour of $s^{(\infty)}$ some traces of the quantum phase transitions of this special state.
%
%

We would like to stress that in all the cases we have considered, $s^{(\infty)}$ is always well below the thermal
$T=\infty$ value $\log(2)$ (being the system periodically driven, due to the absence of energy conservation, it can only thermalize at $T=\infty$~\cite{Rigol_PRX14,Lazarides_PRE14,Ponte_AP15,Rosch_PRA15,Russomanno_EPL15}). 
The system cannot thermalize because it is integrable and locally relaxes to the Floquet Generalized Gibbs ensemble Eq.~\eqref{periodic_GGE:eqn}.
%

As we go to small frequencies, $s^{(\infty)}$ decreases. 
This is expected in the adiabatic limit, where the driving excites a vanishingly small number of excitations: there is indeed nothing which
can propagate quantum correlations. The adiabatic limit occurs when the characteristic frequency scale of the perturbation is much smaller
than the minimum energy gap encountered during the dynamics~\cite{Messiah:book}. 
Thanks to the factorization of the Hamiltonian (see Eq.~\eqref{Ht:eqn}), only the single-particle gaps 
(those for each $2\times 2$ $k-$subspace) are important: the adiabatic limit hence corresponds to  
  $\omega_0\ll 2|h_0-h_c|$.
From a numerical point of view, it would be quite time-consuming to consider very small frequencies in Fig.~\ref{sinfomega:fig}. 
When the frequency is very small, the Floquet spectrum undergoes multiple foldings in the first Floquet-Brillouin zone 
$[-\omega_0/2,\omega_0/2]$: as $k$ changes the quasi-energies $\mu_k^\pm$ undergo many avoided crossings. 
At each crossing, the quantities $|r_k^\pm|^2$ 
undergo quite sudden changes (see the SM of~\cite{Russomanno_PRL12}): hence
the integrand in Eq.~\eqref{sel_leading:eqn} is a strongly oscillating function of $k$. 
For very low frequencies, we need indeed very fine meshes in $k$ (hence, very long chains) to obtain a good estimate of the integral.
%

\section{Conclusion and perspectives} \label{conclusion:sec}
We have discussed the behaviour of the entanglement entropy in a time-periodically-driven quantum Ising chain in 
transverse field. We have observed that the entanglement entropy relaxes asymptotically to a time-periodic value with the same period of the driving.
The asymptotic entropy is attained after a transient in time whose duration scales linearly with the size of the subsystem;
we are able to express the asymptotic regime in terms of the Floquet states of the driven system. 

The asymptotic entropy, at leading order, follows a volume law in the size of the chain:
we have found a semi-analytical formula for the coefficient of the leading order term. 
Quite remarkably, this coefficient is not periodic but time-independent; its dependence on $\omega_0$ is reminiscent 
of quantum phase transitions of the Floquet ground state.

Although following a volume law, the asymptotic entropy stays always well below the thermal
value: the system is integrable and locally relaxes to a Floquet GGE ensemble. 
{Noteworthy, the leading term in the asymptotic entropy
is proportional to the von Neumann entropy of the Floquet GGE density matrix. 
So the Floquet GGE gives the asymptotic properties of the system at any arbitrarily long spatial scale $l$. }

It is interesting to consider what happens to the picture we have described in the presence of disorder. 
In principle, arbitrary inhomogenous couplings can be tackled by the approach presented in \ref{Bogoliubov:sec}.
But disorder should lead to localization of the states, and hence correlations should not be able to propagate 
efficiently~\cite{Russomanno_JSTAT13,Ziraldo_PRL12,art:Ziraldo_PRB13}.
We defer to future publications the analysis of this case and the study of different time-driving protocols.
It would also be interesting to know if an analysis for generic driven one-dimensional systems based on conformal
field theory is possible, in the spirit of what Ref.~\cite{Calabrese_JSTAT05} does for quenched one-dimensional systems.

%
\ack
We acknowledge important discussions with E.~G. Dalla Torre, A. De Pasquale and J. Goold. Research was supported by  
the EU FP7 under grant agreements n.~280555,  n.~600645 (IP-SIQS), n.~618074 (STREP-TERMIQ), n.~641122 (STREP-QUIC), 
by the Coleman-Soref foundation, by the Israeli Science Foundation (grant number $1542/14$), by UNICREDIT and by National Research Foundation, Prime Minister's Office, Singapore under its Competitive Research Programme (CRP-QSYNC Award No. NRF- CRP14-2014-02). This work was not supported
by any military agency.
\appendix
%

\section{- Bogoliubov-de Gennes dynamics} \label{Bogoliubov:sec}
%
In this Appendix we briefly describe the quantum dynamics of Ising/XY chains~\cite{Caneva_PRB07,Russomanno_JSTAT13}. 
Here we do not assume translational invariance, therefore Fourier transform to momentum space cannot be employed.
We closely follow the discussion of Ref.~\cite{Russomanno_JSTAT13}. 
Generically, if $\opc{j}$ denote the $L$ fermionic operators originating from the Jordan-Wigner transformation of spin operators~\cite{Lieb_AP61}
\begin{eqnarray}
  \label{wj}
  \hat{\sigma}_j^x&=&1-2\opcdag{j} \opc{j}\nonumber\\
  \hat{\sigma}_j^z&=&\tau_j\left(\opcdag{j}+\opc{j}\right)\nonumber\\
  \hat{\sigma}_j^y&=&i\tau_j\left(\opcdag{j}-\opc{j}\right)\,,\quad{\rm with}\quad \hat{\tau}_j\equiv\prod_{l<j}\hat{\sigma}_{l}^x\,,
\end{eqnarray}
we can write the Hamiltonian in Eq.~\eqref{h11} as a quadratic fermionic form
\begin{equation} \label{hamor}
 \hat{H}(t) = 
\hat{\mathbf{\Psi}}^{\dagger} \cdot {\mathbb H}(t) \cdot \hat{\mathbf{\Psi}}  
  = \left( \begin{array}{cc}  \opbfcdag{} & \opbfc{} \end{array} \right)
  \left( \begin{array}{rr} {\bf A}(t) & {\bf B}(t) \\
                                        -{\bf B}(t) & -{\bf A}(t) \end{array} \right)
                                        \left( \begin{array}{l}  \opbfc{} \\ \opbfcdag{} \end{array} \right)
                                        \;.
\end{equation}
Here $\hat{\mathbf{\Psi}}$ are $2L$-components (Nambu) fermionic operators defined as 
$\hat{\Psi}_j = \hat{c}_j$  (for $1\le j\le L$) and $\hat{\Psi}_{L+j} = \hat{c}_j^\dagger$,  
and  ${\mathbb H}$ is a $2L\times 2L$ Hermitian matrix having the explicit form shown on the right-hand side, 
with $\bf A$ an $L\times L$ real symmetric matrix, $\bf B$ an $L\times L$ real anti-symmetric matrix. 
Such a form of ${\mathbb H}$ implies a particle-hole symmetry: if $({\bf u}_{\alpha}, {\bf v}_{\alpha})^T$ is an instantaneous eigenvector 
of ${\mathbb H}$ with eigenvalue $\epsilon_{\alpha}\ge 0$, then $({\bf v}_{\alpha}^*, {\bf u}_{\alpha}^*)^T$ is an eigenvector with 
eigenvalue $-\epsilon_{\alpha}\le 0$.

Let us now focus on a given time, $t=0$, or alternatively suppose that the Hamiltonian is time-independent. 
Then, we can apply a unitary Bogoliubov transformation  
\begin{equation} \label{trasgro}
  \hat{\mathbf{\Psi}} = \left( \begin{array}{l}  \opbfc{} \\ \opbfcdag{} \end{array} \right) = 
  {\mathbb U}_0 \cdot \left( \begin{array}{l} \opbfgamma{} \\ \opbfgammadag{} \end{array} \right) =
  \left(\begin{array}{rr} {\bf U}_0 & {\bf V}^*_0 \\
                                       {\bf V}_0 & {\bf U}^*_0 \end{array} \right) \cdot  
                                       \left( \begin{array}{l} \opbfgamma{} \\ \opbfgammadag{} \end{array} \right) \;,
\end{equation}
where ${\bf U}_0$ and ${\bf V}_0$ are $L\times L$ matrices collecting all the eigenvectors of $\mathbb H$, by column, turning the Hamiltonian 
in Eq.~\eqref{hamor} in the diagonal form
\begin{equation}
  \hat{H} = \sum_{\alpha=1}^L \epsilon_\alpha \left(\opgammadag{\alpha} \opgamma{\alpha} - \opgamma{\alpha}\opgammadag{\alpha}\right)\,,
\end{equation}
where the $\opgamma{\alpha}$ are new quasiparticle Fermionic operators. 
%
%
The ground state $\ket{{\rm GS}}$ has energy $E_{\rm GS}=-\sum_\alpha \epsilon_\alpha$ and is the vacuum of the $\opgamma{\alpha}$ for all values of 
$\alpha$: $\opgamma{\alpha}\ket{{\rm GS}}=0$. 

To discuss the quantum dynamics when ${\hat H}(t)$ depends on time, one starts by writing the Heisenberg equations of
motion for the $\hat{\mathbf{\Psi}}$, which turn out to be {\em linear}, due to the quadratic nature of $\hat{H}(t)$.
A simple calculation shows that:
\begin{equation}
i\hbar \frac{d}{dt} \hat{\mathbf{\Psi}}_{H}(t) = 2 {\mathbb H}(t) \cdot \hat{\mathbf{\Psi}}_{H}(t) \;,
\end{equation}
the factor $2$ on the right-hand side originating from the off-diagonal contributions due to $\{\Psi_j,\Psi_{L+j}\}=1$.
These Heisenberg equations should be solved with the initial condition that, at time $t=0$, is
\begin{equation} \label{trasgro_0}
  \hat{\mathbf{\Psi}}_H(t=0) = \hat{\mathbf{\Psi}} = 
{\mathbb U}_0 \cdot  \left( \begin{array}{l} \opbfgamma{} \\ \opbfgammadag{} \end{array} \right) \;.
\end{equation}
A solution is evidently given by 
\begin{equation} \label{Psi-Heis:eqn}
  \hat{\mathbf{\Psi}}_H(t) = 
{\mathbb U}(t) \cdot  \left( \begin{array}{l} \opbfgamma{} \\ \opbfgammadag{} \end{array} \right) =
\left(\begin{array}{rr} {\bf U}(t) & {\bf V}^*(t) \\
                                       {\bf V}(t) & {\bf U}^*(t) \end{array} \right) \cdot  
                                       \left( \begin{array}{l} \opbfgamma{} \\ \opbfgammadag{} \end{array} \right)
\end{equation}
with the same $\opbfgamma{}$ used to diagonalize the initial $t=0$ problem, as long as the time-dependent coefficients
of the unitary $2L\times2L$ matrix ${\mathbb U}(t)$ satisfy the ordinary linear Bogoliubov-de Gennes time-dependent equations:
\begin{equation} \label{bog}
i\hbar \frac{d}{dt} {\mathbb U}(t) 
= 2{\mathbb H} (t) \cdot 
{\mathbb U}(t) 
\end{equation}
with initial conditions ${\mathbb U}(t=0)={\mathbb U}_0$. 
It is easy to verify that the time-dependent Bogoliubov-de Gennes form implies that the operators $\opgamma{\alpha}(t)$ in 
the Schr\"odinger picture are time-dependent and annihilate the time-dependent state $|\psi(t)\rangle$.
Notice that ${\mathbb U}(t)$ looks like the unitary evolution operator of a $2L$-dimensional problem with Hamiltonian $2{\mathbb H}(t)$.
This implies that one can use a Floquet analysis to get ${\mathbb U}(t)$ whenever ${\mathbb H}(t)$ is time-periodic. 
This trick provides us with {\em single-particle} Floquet modes and quasi-energies in terms of which we can evaluate,
through the Heisenberg picture prescription, any two-point correlator $\mean{\opc{i}\opc{j}}$, $\mean{\opcdag{i}\opc{j}}$. 
Because the state is Gaussian, from those objects one can reconstruct the expectation of any operator and also the entanglement
entropy of any subchain as we show in~\ref{enta_entra:sec}. To that purpose, it is very useful to rewrite Eq.~\eqref{Psi-Heis:eqn}
in the Schr\"odinger representation: we obtain
\begin{equation} \label{Psi-Schro:eqn}
  \left( \begin{array}{l} \opbfc{} \\ \opbfcdag{} \end{array} \right) = 
\left(\begin{array}{rr} {\bf U}(t) & {\bf V}^*(t) \\
                                       {\bf V}(t) & {\bf U}^*(t) \end{array} \right)
\cdot  \left( \begin{array}{l} \opbfgamma{}(t) \\ \opbfgammadag{}(t) \end{array} \right)\,.
\end{equation}
Here we have defined $\opbfgamma{}(t)$ as an  $L$-components fermionic operators defined as 
$\opgamma{\mu}(t)=\hat{U}(t)\opgamma{\mu}\hat{U}^\dagger(t)$ for $\mu=1,\ldots,L$; $\hat{U}(t)$ is the
time-evolution operator in the full Hilbert space such that $\hat{U}(0)=\hat{\boldsymbol{1}}$ and 
$i\hbar\partial_t\hat{U}(t)=\hat{H}(t)\hat{U}(t)$. With these definitions, it is easy to see that
\begin{equation} \label{def-destroy}
  \opgamma{\mu}(t)\ket{\psi(t)} = 0\quad\forall\mu,\;\quad\forall t\,.
\end{equation}

%
\section{- Evaluation of the entanglement entropy} \label{enta_entra:sec}
In this section we show explicitly the computation of the entanglement entropy in the driven quantum Ising chain which leads
to the results of Sec.~\ref{enta_dyna:sec}.
We express the entropy in terms of the solution of the dynamical Bogoliubov-de Gennes equations (\ref{Bogoliubov:sec}). 
We generalize to the case of driven
systems the methods for the evaluation of the entanglement entropy in the static chain introduced in Ref.~\cite{Latorre_QIC04}.
%
\subsection{- Correlation matrix of the Majorana operators} \label{generic_corr:sec}
In this subsection we introduce Majorana fermions starting from the Bogoliubov fermions of~\ref{Bogoliubov:sec} and
then we express their correlation matrix in terms of the solution of the dynamical Bogoliubov-de Gennes equations.
We are going to see in Subsection~\ref{entroval_subsec} that this correlation matrix is an essential ingredient 
to evaluate the entanglement entropy of any sub-chain. We are going to see first the generic case and then the
translationally-invariant one. In the latter case, the correlation matrix assumes a Block-diagonal form in $k$-space, which
allows to fasten the numerical calculations.
\subsubsection*{Generic, non translationally invariant case.}
We consider a generic form of external driving: we can apply the results of~\ref{Bogoliubov:sec}. In particular, by inverting
the unitary matrix $\mathbb{U}(t)$ obeying Eq.~\eqref{bog}, we can rewrite Eq.~\eqref{Psi-Schro:eqn} in the form
\begin{equation} \label{transfor}
  \left(\begin{array}{c}\opgamma{\mu}(t)\\
                        \opgammadag{\mu}(t)\end{array}\right)= 
      \displaystyle\sum^L_{j=1} \left[\begin{array}{cc}U_{j{\mu}}^*(t) &V_{j{\mu}}^*(t)\\
                                                       V_{j{\mu}}(t)   &U_{j{\mu}}(t)\end{array}\right]
                                \left(\begin{array}{c}\opc{j}\\
                                                     \opcdag{j}
                                      \end{array}\right)\,.
\end{equation}
%
To obtain the entanglement entropy,
we have to introduce two sets of Majorana operators (see Refs.~\cite{Latorre_QIC04,Alicea_RPP12,Wilczek_NP09})
\begin{eqnarray} \label{transmaj}
%
  \left(\begin{array}{c}\mopc{2l-1}\\
                        \mopc{2l}\end{array}\right)&=&
  \left[\begin{array}{cc}1&1\\
                        -i&i\end{array}\right]  
   \left(\begin{array}{c}\opc{l}\\\opcdag{l}\end{array}\right)
%
\quad{\rm for}\quad l=1,\ldots,L\quad{\rm and}\nonumber\\
  \left(\begin{array}{c}\mopgamma{2\mu-1}(t)\\
                        \mopgamma{2\mu}(t)\end{array}\right)&=&
  \left[\begin{array}{cc}1&1\\
                        -i&i\end{array}\right]  
   \left(\begin{array}{c}\opgamma{\mu}(t)\\
                         \opgammadag{\mu}(t)\end{array}\right)
%
%
\quad{\rm for}\quad \mu=1,\ldots,L\,.
\end{eqnarray}
It is easy to see from the definitions that these objects obey the Majorana algebra: 
$\left\{\mopgamma{\alpha},\,\mopgamma{\beta}\right\}=\delta_{\alpha\beta}$ and $\mopgamma{\alpha}=\mopgammadag{\alpha}$ $\forall\alpha,\,\beta$.
Applying those transformations to Eq.~\eqref{transfor} we can immediately find
\begin{equation}  \label{diagmaj1}
  \left(\begin{array}{c}
   \mopgamma{2\mu-1}\\
   \mopgamma{2\mu}
  \end{array}\right) =
  \displaystyle\sum^L_{j=1}\left[\begin{array}{cc}
   \Real U_{j{\mu}}+\Real V_{j{\mu}}&\Aimag U_{j{\mu}}-\Aimag V_{j{\mu}}\\
   -\left(\Aimag U_{j{\mu}}+\Aimag V_{j{\mu}}\right)&\Real U_{j{\mu}}-\Real V_{j{\mu}}
  \end{array}\right]
  \left(\begin{array}{c}
   \mopc{2j-1}\\
   \mopc{2j}
  \end{array}\right)\,.
\end{equation}
We can recast this formula as
\begin{equation}  \label{diagmaj2}
  \mopgamma{\sigma}(t)=\sum_{n=1}^L W_{\sigma n}(t)\; \mopc{n}\,.
\end{equation}
Thanks to the unitarity of the $L\times L$ matrix $\mathbb U^\dagger(t)=\left( \begin{array}{cc} \U^\dagger(t)&\V^\dagger(t)\\ \V^T(t)&\U^T(t) \end{array} \right)$, 
we can see that the $2L\times 2L$ matrix $\mathbb W(t)$ ($[\mathbb W(t)]_{\sigma n}=W_{\sigma n}(t)$) is an 
orthogonal matrix $\mathbb W^{-1}(t)=\mathbb W^T(t)$\,.
At this point we can evaluate the correlation matrix of the operators $\mopc{l}$. To do that, we notice that --- thanks to Eq.~\eqref{def-destroy} --- we have the
following averages
\begin{eqnarray}
  \mean{\opgamma{\mu}(t)}_t&=&0\nonumber\\
  \mean{\opgamma{\mu}(t)\opgamma{\nu}(t)}_t&=&0\nonumber\\
  \mean{\opgamma{\mu}(t)\opgammadag{\nu}(t)}_t&=&\delta_{\mu\nu}\,,
\end{eqnarray}
where the bracket indicates the average over the state $\ket{\psi(t)}$. Using Eq.~\eqref{transmaj} it is easy to verify that
\footnote
{The explicit calculation, performed using the anticommutation properties of the Majorana operators, is
$$
1=\mean{\opgamma{\mu}\opgammadag{\mu}}_t=\frac{1}{4}\mean{(\mopgamma{2\mu-1}+i\mopgamma{2\mu})(\mopgamma{2\mu-1}-i\mopgamma{2\mu})}_t=
\frac{1}{4}\left(2-i\mean{\left[\mopgamma{2\mu-1},\mopgamma{2\mu}\right]}_t\right)=\frac{1}{2}\left[1-i\mean{\mopgamma{2\mu-1}\mopgamma{2\mu}}_t\right]\,.
$$
In this way we show that $\mean{\mopgamma{2\mu-1}\mopgamma{2\mu}}_t=-\mean{\mopgamma{2\mu}\mopgamma{2\mu-1}}_t=i$. Because of the anticommutation
rules, we have also $\mean{(\mopgamma{2\mu-1})^2}_t=\mean{(\mopgamma{2\mu})^2}_t=1$. To show that all the other correlators of those
Majorana operators vanish, we take $\nu\neq\mu$ and the conclusion follows substituting Eq.~\eqref{transmaj} in the relations
$\mean{\opgamma\mu\opgamma\nu}_t=\mean{\opgamma\mu\opgammadag\nu}_t=\mean{\opgamma\nu\opgammadag\nu}_t=0$.
}
 the correlation matrix is
\begin{equation} \label{mat_corr_diag}
  \mean{\mopgamma{\sigma}(t)\mopgamma{\eta}(t)}_t=\delta_{\sigma\eta}+i\Gamma_{L\,\sigma{\eta}}^{\,\gamma}
\end{equation}
where we have defined the $2L\times 2L$ matrix
$$
\boldsymbol{ \Gamma}_{L}^{\,\gamma} = \bigoplus_{\mu=1}^L
  \left[\begin{array}{cc}0&1\\
       -1&0\end{array}\right]\,.
$$
The correlation matrix of the original Majorana operators $\mopc{n}$ 
can be obtained from this one by means of the transformation Eq.~\eqref{diagmaj2}; the orthogonality of the transformation
matrix $\mathbb W$ and Eq.~\eqref{mat_corr_diag} give
\begin{equation} \label{correl_mat}
  \mean{\mopc{m}\mopc{n}}_t=\delta_{m{n}}+i\Gamma_{L\,m{n}}^{\,C}(t)\,;
\end{equation}
where we have defined the $2L\times 2L$ matrix 
\begin{equation} \label{gener_corr:eqn}
  \boldsymbol{\Gamma}_{L}^{\,C}(t)\equiv\mathbb W^T(t)\,\boldsymbol{\Gamma}_{L}^{\,\gamma}\,\mathbb W(t)\,.
\end{equation}
Both the matrices $\boldsymbol{\Gamma}_{L}^{\,C}(t)$ and $\boldsymbol{\Gamma}_{L}^{\,\gamma}$ are skew-symmetric. 
The coefficients $U_{j\mu}$ and $V_{j\mu}$
from which these matrices are obtained can be numerically evaluated by solving the Bogoliubov-de Gennes equations Eq.~\eqref{bog}. 
\subsubsection*{Translationally invariant case.}
When the system is translationally invariant,
it is possible to relate the coefficients $U_{j{\mu}}$, $V_{j{\mu}}$ of the matrix $\mathbb{W}(t)$ to the coefficients $u_k$
and $v_k$, rapidly found by moving to $k$-space and integrating there 
the $L$ decoupled $2\times 2$ Bogoliubov-de Gennes systems of equations Eq.~\eqref{deGennes:eqn}. 
Here the Fermionic Hamiltonian is block-diagonal (see Eq.~\eqref{Ht:eqn}).
%
%
%
Also the Bogoliubov operators Eq.~\eqref{def-destroy} have definite momentum and are related to the $\opc{k}$ by means of a unitary transformation
constructed by means of the $u_k(t)$ and $v_k(t)$ of Eq.~\eqref{state:eqn}
\begin{equation}
  \left(\begin{array}{c}
                        \opgamma{k}(t)\\ \opgammadag{-k}(t)
                \end{array} \right) = 
               \left(\begin{array}{cc}
                        u_k^*(t)&v_k^*(t)\\
                        -v_k(t)&u_k(t)
                \end{array}        \right)\cdot
                 \left(\begin{array}{c}
                        \opc{k}\\ \opcdag{-k}
                \end{array} \right)\,.
\end{equation}
Introducing the Majorana operators as in Eq.~\eqref{transmaj}, we find that the $\mopgamma{2k-1}$ and the $\mopc{2k-1}$ are related
by means of the orthogonal transformation
\begin{equation}
    \left(\begin{array}{c}
                        \mopgamma{2k-1}(t)\\ 
                        \mopgamma{2k}(t)\\
                        \mopgamma{-2k-1}(t)\\
                        \mopgamma{-2k}(t)
                \end{array} \right) = {\mathbb M}_k
                 \left(\begin{array}{c}
                        \mopc{2k-1}\\ 
                       \mopc{2k}\\
                        \mopc{-2k-1}\\
                       \mopc{-2k}
                \end{array} \right){\rm ,}\, 
                \mathbb{M}_k\equiv\left(\begin{array}{cc|cc}
                        \Real u_k(t)&\Aimag u_k(t)&\Real v_k(t)&-\Aimag v_k(t)\\
                        -\Aimag u_k(t)&\Real u_k(t)&-\Aimag v_k(t) & -\Real v_k(t)\\
                        \hline
                        -\Real v_k(t) & \Aimag v_k(t) & \Real u_k(t) & \Aimag u_k(t)\\
                        \Aimag v_k(t) & \Real v_k(t) & -\Aimag u_k(t) & \Real u_k(t)
                \end{array}        \right).
\end{equation}
Using the polar representation $u_k(t)=|u_k(t)|\nep^{i\theta_k(t)}$, $v_k(t)=|v_k(t)|\nep^{i\varphi_k(t)}$,
we can write the matrix $\mathbb{M}_k(t)$ in the form
\begin{displaymath}
  \mathbb{M}_k(t) = \left(\begin{array}{c|c}{\bf U}_k&{\bf V}_k\\
                     \hline
                     -{\bf V}_k&{\bf U}_k\end{array}\right)
   \quad{\rm with}\quad{\bf V}_k=|v_k(t)|\sigma^z\nep^{-i\varphi_k(t)\sigma^y}={\bf V}_k^T\quad{\rm and}
    \quad{\bf U}_k=|u_k(t)|\nep^{i\theta_k(t)\sigma^y}\,.
\end{displaymath}
Because of the orthogonality of the matrix $\mathbb{M}_k(t)$, we have the relations
\begin{eqnarray}
  {\bf U}_k^T{\bf U}_k + {\bf V}_k^T{\bf V}_k&=&\boldsymbol{1}\nonumber\\
  {\bf U}_k{\bf U}_k + {\bf V}_k{\bf V}_k&=&\boldsymbol{0}\,.
\end{eqnarray}
Thanks to Eq.~\eqref{def-destroy}, we see that the correlation of the $\mopgamma{2k-1}$ has the $k$-factorized form
\begin{displaymath}
  \boldsymbol{1}_{2L} +i\boldsymbol{\Gamma}^\gamma_{L}\equiv\bigoplus_{k>0}\mean{ \left(\begin{array}{c}
                        \mopgamma{2k-1}\\ 
                        \mopgamma{2k}\\
                        \mopgamma{-2k-1}\\
                        \mopgamma{-2k}
                \end{array} \right)\otimes  \left(\begin{array}{cccc}
                        \mopgamma{2k-1}&
                        \mopgamma{2k}&
                        \mopgamma{-2k-1}&
                        \mopgamma{-2k}
                \end{array} \right)} = \boldsymbol{1}_{2L} \,-\,
                  \bigoplus_{k>0}\left(\begin{array}{c|c}
                         \sigma^y&\boldsymbol{0}\\
                       \hline
                     \boldsymbol{0}&\sigma^y
                    \end{array}\right)\,.
\end{displaymath}
From this we can find the correlation matrix of the $\opc{k}$ operators, defined analogously, which is still factorized in $k$:
it has the form
\begin{eqnarray}
  \boldsymbol{1}_{2L} +i\boldsymbol{\Gamma}^{C_k}_{L}(t) &=&\boldsymbol{1}_{2L} 
    \,-\,\bigoplus_{k>0} \mathbb{M}_k^T(t) \left(\begin{array}{c|c}
                         \sigma^y&\boldsymbol{0}\\
                       \hline
                     \boldsymbol{0}&\sigma^y
                    \end{array}\right)
            \mathbb{M}_k(t)\\
          &=&\boldsymbol{1}_{2L} - \bigoplus_{k>0}\left(\begin{array}{c|c}
             {\bf U}_k^T\sigma^y{\bf U}_k+{\bf V}_k\sigma^y{\bf V}_k & 
                     {\bf U}_k^T\sigma^y{\bf V}_k-{\bf V}_k\sigma^y{\bf U}_k\\
                     \hline
              {\bf V}_k\sigma^y{\bf U}_k-{\bf U}_k^T\sigma^y{\bf V}_k & 
                     {\bf U}_k^T\sigma^y{\bf U}_k+{\bf V}_k\sigma^y{\bf V}_k\end{array}\right)\,.\nonumber
\end{eqnarray}
It is not difficult to see that
\begin{eqnarray}
  &&\hspace{-1cm}{\bf U}_k^T\sigma^y{\bf U}_k+{\bf V}_k\sigma^y{\bf V}_k=\left(1-2|v_k(t)|^2\right)\sigma^y\,,\nonumber\\
  &&\hspace{-1cm}{\bf U}_k^T\sigma^y{\bf V}_k-{\bf V}_k\sigma^y{\bf U}_k=
   2i|u_k(t)||v_k(t)|\sigma^x\nep^{i[\theta_k(t)-\varphi_k(t)]\sigma^y}\nonumber\\
   &&\hspace{1.5cm}=2i\left[\Real(u_k(t)v_k^*(t))\,\sigma^x-\Aimag(u_k(t)v_k^*(t))\,\sigma^z\right]\,.
\end{eqnarray}
From this we can find the correlation matrix $\boldsymbol{1}_{2L} +\Gamma^{C}_{L}(t)$ of the operators $\mopc{2j-1}$. Because we have
\begin{displaymath}
  \opc{j} = \frac{1}{\sqrt{L}}\sum_{k}\nep^{-ikj}\opc{k}\,,
\end{displaymath}
we can write
\begin{displaymath}
  \left(\begin{array}{c}
                        \mopc{2j-1}\\ 
                        \mopc{2j}
                \end{array} \right) = \frac{1}{\sqrt{L}}\sum_k \left(\begin{array}{cc}
                                             \cos(kj) & \sin(kj) \\
                                             -\sin(kj) & \cos(k j)
                                            \end{array}\right)\cdot
                                       \left(\begin{array}{c}
                                    \mopc{2k-1}\\ 
                                    \mopc{2k}
                                      \end{array} \right)=
                               \frac{1}{\sqrt{L}}\sum_k\nep^{i(kj)\sigma^y}\left(\begin{array}{c}
                                    \mopc{2k-1}\\ 
                                    \mopc{2k}
                                      \end{array} \right)\,.
\end{displaymath}
From this relation, defining the matrix
\begin{displaymath}
  \mathbb{Q}_{k,j} = \left(\begin{array}{c|c}
                         \nep^{i(kj)\sigma^y}&\nep^{-i(kj)\sigma^y}
                      \end{array}\right)\,,
\end{displaymath}
we find that the correlation matrix is a $2L\times 2L$ matrix composed by $2\times 2$ blocks. The block positioned in the row $j$ and the
column $l$ has the form
\begin{eqnarray} \label{matriciona:eqn}
  &&\hspace{-2.5cm}\boldsymbol{1}_{2}\,\delta_{jl} + i{\Gamma}^{C}_{L\,jl}(t) = 
       \mean{\left(\begin{array}{c}
                     \mopc{2j-1}\\
                     \mopc{2j}\end{array}\right)\otimes
                      \left(\begin{array}{cc}
                     \mopc{2l-1}&
                     \mopc{2l}\end{array}\right)} = \boldsymbol{1}_{2}\,\delta_{jl}-\bigoplus_{k>0}
                    \mathbb{Q}_{k,j}\mathbb{M}_k^T(t)\left(\begin{array}{c|c}
                         \sigma^y&\boldsymbol{0}\\
                       \hline
                     \boldsymbol{0}&\sigma^y
                    \end{array}\right)\mathbb{M}_k(t)\mathbb{Q}_{k,l}^T\nonumber\\
  &\quad&\nonumber\\
  &&\hspace{-2cm}=\boldsymbol{1}_{2}\,\delta_{jl}- \frac{1}{L}\sum_{k>0} 
            \left(\begin{array}{c|c}
                         \nep^{i(kj)\sigma^y}&\nep^{-i(kj)\sigma^y}
                      \end{array}\right)\nonumber\\
   &&\hspace{-1.5cm} \times\left(\begin{array}{c|c}
             \left(1-2|v_k(t)|^2\right)\sigma^y & {\small 2i|u_k(t)||v_k(t)|\sigma^x\nep^{i[\theta_k(t)-\varphi_k(t)]\sigma^y}}\\
                     \hline
              {\small-2i|u_k(t)||v_k(t)|\sigma^x\nep^{i[\theta_k(t)-\varphi_k(t)]\sigma^y}} & \left(1-2|v_k(t)|^2\right)\sigma^y\end{array}\right)
              \left(\begin{array}{c}
                         \nep^{-i(kl)\sigma^y}\\\hline\nep^{i(kl)\sigma^y}
                      \end{array}\right)\nonumber\\
  &\quad&\nonumber\\
   &&\hspace{-2.cm}=\boldsymbol{1}_{2}\delta_{jl} - \frac{2}{L}\sum_{k>0} \Big[[|u_k(t)|^2-|v_k(t)|^2]\,\sigma^y\cos(k(j-l))
       \nonumber\\
   &&\hspace{-1.5cm}-2i(\Real(u_k(t)v_k^*(t))\,\sigma^z+\Aimag(u_k(t)v_k^*(t))\,\sigma^x)\sin(k(j-l))\Big]\,.\nonumber\\
\end{eqnarray}
%
%
%
%
%
%
Let us go into the thermodynamic limit, where the discrete summation becomes an integral. Let us also consider -- for reasons
that will be clear soon -- the $2l\times 2l$ submatrix $\boldsymbol{\Gamma}_{l}^C$ which corresponds to some specific subsystem.
This submatrix has the block Toeplitz form 
%
\begin{equation} \label{corr_mat_tras_t:eqn}
  \Gamma_l^C(t)=\left(\begin{array}{cccc}\Pi_0(t)&\Pi_{-1}(t)&\cdots&\Pi_{1-l}(t)\\
                                      \Pi_1(t)&\Pi_0(t)&&\vdots\\
                                      \vdots&&\ddots&\vdots\\
                                      \Pi_{l-1}(t)&\cdots&\cdots&\Pi_0(t)\end{array}\right),\quad 
    \Pi_l(t)=\left(\begin{array}{cc}-2i{R}_l(t)&{Q}_l(t)-2iI_l(t)\\
                                 -\left({Q}_{-l}(t)-2i I_{-l}(t)\right)&2i{R}_l(t)\end{array}\right)\,,
\end{equation}
where we have defined
\begin{eqnarray} \label{R_cacca:eqn}
  R_l(t)&=&\frac{1}{2\pi}\int_{-\pi}^\pi \nep^{-ikl} R_k(t)\,\ud k,\quad R_k(t)\equiv \Real\left(u_k(t)v_k^*(t)\right)\\
  I_l(t)&=&\frac{1}{2\pi}\int_{-\pi}^\pi \nep^{-ikl} I_k(t)\,\ud k,\quad I_k(t)\equiv \Aimag\left(u_k(t)v_k^*(t)\right)\\
  Q_l(t)&=&\frac{1}{2\pi}\int_{-\pi}^\pi \nep^{-ikl} Q_k(t)\,\ud k,\quad Q_k(t)\equiv \left|u_k(t)\right|^2-\left|v_k(t)\right|^2\,.
\end{eqnarray}
The $u_k(t)$ and $v_k(t)$ are solutions of Eq.~\eqref{deGennes:eqn}
\footnote{We note in passing that the  
vector $\bf{B}_k(t)\equiv\left(\begin{array}{c}2R_k(t)\\2I_k(t)\\Q_k(t)\end{array}\right)$ is the Bloch-sphere representative of
the state $\ket{\psi_k(t)}=\left(\begin{array}{c}u_k(t)\\v_k(t)\end{array}\right)$. Moreover, it is not difficult to verify that --
taking for $u_k$ and $v_k$ the values which these objects assume in the ground state -- Eq.~\eqref{corr_mat_tras_t:eqn}
reduces to the well-known formula given in Ref.~\cite{Latorre_QIC04}}.
The Hamiltonian matrix in Eq.~\eqref{deGennes:eqn}
obeys the symmetry relation $\mathbb{H}_{-k}(t)=\sigma_z\mathbb{H}_{k}(t)\sigma_z$: this implies that $u_k(t)=u_{-k}(t)$ 
and $v_k(t)=-v_{-k}(t)$. From this it follows that $R_{-l}=-R_l$, $I_{-l}=-I_l$, $Q_{-l}=Q_l$: this implies that the matrix $\Gamma_l^C(t)$
is antisymmetric, as it should be. It is easy to see that the matrix $i\Gamma_l^C(t)$ is Hermitian: it has $2l$ eigenvalues of the 
form $\pm\nu_m(t),\quad m=1,\ldots l$. In the next subsection we are going to show how to use these eigenvalues to evaluate the entanglement
entropy of the considered subchain.
\subsection{ - Entanglement entropy}  \label{entroval_subsec}
From the correlation matrix (Eq.~\eqref{gener_corr:eqn} or Eq.~\eqref{corr_mat_tras_t:eqn}), the entanglement entropy of any subchain can be evaluated by
means of the standard method introduced in Ref.~\cite{Latorre_QIC04}. For reader convenience, we report this discussion in the present subsection.
We  select a subchain made by $l$ adjacent spins. 
The entanglement entropy associated to this partition is evaluated as the Von Neumann entropy 
of the reduced density matrix $\hat{\rho}_l$ for the $l$ adjacent spins
$$
  S_l(t) = -\Tr\left[\hat{\rho}_l(t)\log\hat{\rho}_l(t)\right]\,.
$$
%
%
%
We can expand the density matrix $\rho_l$ of the block as
\begin{equation} \label{expansion}
  \hat{\rho}_l(t)=2^{-l}\sum_{\mu_1,\,\ldots,\,\mu_l=0,x,y,z}\rho_{\mu_1\cdots\mu_l}(t)\,\hat{\sigma}_1^{\mu_1}\cdots\hat{\sigma}_l^{\mu_l}\,,
\end{equation}
(we define $\hat{\sigma}^0=\boldsymbol{1}_2$) where the time-dependent coefficients $\rho_{\mu_1\cdots\mu_l}(t)$ are given by
\begin{equation} \label{average:eqn}
  \rho_{\mu_1\cdots\mu_l}(t) = \mean{\hat{\sigma}_1^{\mu_1}\cdots\hat{\sigma}_l^{\mu_l}}_t\,.
\end{equation}
The Hamiltonian is invariant under Fermionic parity transformation
\begin{equation}  \label{parinvaria}
  \left(\prod_{j=1}^L\hat{\sigma}_{j}^x\right)\hat{H}(t)\left(\prod_{j=1}^L\hat{\sigma}_j^x\right)=\hat{H}(t)\quad\forall t\,.
\end{equation}
As they show in~\cite{Latorre_QIC04}, this implies that $\rho_{\mu_1\cdots\mu_l}(t)=0$ whenever the number of indices $j$ for
which $\mu=y$ and $\mu=z$ is odd.

Applying the Jordan-Wigner transformation, therefore, the string operators $\tau_j$ (see Eq.~\eqref{wj}) appear always in pairs: their tails (which
are infinitely long in the thermodynamic limit) annihilate each other. Hence, we can express the expectation in Eq.~\eqref{average:eqn} as an algebraic
sum of many-point Fermionic
correlators, each containing a finite number of creation/annihilation operators. Because the state of the system is Gaussian, 
we can apply Wick theorem and express the many-point correlators as sums of products of the two-point Majorana
correlators forming the subsystem correlation matrix $\boldsymbol{1}_{2l} +i\boldsymbol{\Gamma}_l^C(t)$
(see Eq.~\eqref{corr_mat_tras_t:eqn}). 

As we are going to describe in detail, we can block-diagonalize this matrix and show that the considered subsystem is equivalent
to $l$ uncorrelated fermionic modes $\opd{j}(t)$ occupied with probabilities $p_j(t)<1$. The density matrix of the subsytem is the tensor product of the density
matrices of the fermionic modes: each one can be seen as a two-level quantum system where the state $\ket{0}$ is occupied with probability $1-p_j(t)$
and the state $\opddag{j}(t)\ket{0}$ is occupied with probability $p_j(t)$. The entanglement entropy is indeed the sum of the von Neumann entropies of the factorized
fermionic modes: $S_l(t)=-\sum_j=1^l\left[p_j(t)\log p_j(t)+(1-p_j(t))\log(1-p_j(t))\right]$.

Being more precise, we can express the matrix $\boldsymbol{\Gamma}_l^C(t)$
in a block-diagonal form
matrix  
\begin{equation} \label{gammaL:eqn}
  \boldsymbol{\Gamma}_{l}^{\,D}(t)=\mathbb V(t)\,\boldsymbol{\Gamma}_{l}^{\,C}(t)\,\mathbb V^T(t)
\end{equation}
where $\mathbb V(t)$ is a $2l\times 2l$ orthogonal matrix ($\mathbb V^{-1}(t)=\mathbb{V}^T(t)$) and
$\boldsymbol{\Gamma}_{l}^{\,D}(t)$ is written in terms of its eigenvalues
\begin{equation} \label{gamma_l}
\boldsymbol {\Gamma}_{l}^{\,D}(t) = \bigoplus_{j=1}^{l}
  \left[\begin{array}{cc}0&\nu_j(t)\\
       -\nu_j(t)&0\end{array}\right]\,
\end{equation}
(from the technical point
of view, we find the eigenvalues $\nu_j(t)$ by numerically diagonalizing 
the Hermitian matrix $i\boldsymbol{\Gamma}_{2l}^{\,D}(t)$). We can now define the Majorana modes
$$
  \check{d}_p(t)=\sum_{m=1}^{2L}V_{pm}(t)\check{c}_m\,.
$$
From Eq.~\eqref{gammaL:eqn}, we can see that the mode $\mopd{2j-1}$ is only correlated to the mode $\mopd{2j}$:
a most convenient fact that we are going to exploit. 
At this point we can define the $L$ spinless fermionic operators
\begin{eqnarray}
  &&\opd{j}(t)=\frac{1}{2}\left[\mopd{2j-1}(t)+i\mopd{2j}(t)\right]\nonumber\\
  &&\left\{\opd{n}(t),\,\opd{m}(t)\right\}=0,\quad\left\{\opd{n}(t),\,\opddag{m}(t)\right\}=\delta_{n{m}}\,.
\end{eqnarray}
By construction, they fulfill
$$
  \mean{\opd{n}(t)\opd{m}(t)}=0\,\quad \mean{\opddag{n}(t)\opd{m}(t)}=\delta_{n{m}}\frac{1+\nu_m(t)}{2}\,.
$$
This means that the $L$ fermionic modes are uncorrelated: they are in a product state
$$
  \rho_l(t)=\overline{\rho}_1(t)\otimes\ldots\otimes\overline{\rho}_l(t)\,.
$$
This tensor product structure does not correspond in general to a factorization into local Hilbert spaces of the $l$ spins, but is instead a rather non-local
structure. The density matrix $\overline{\rho}_j$ has eigenvalues
\begin{equation} \label{eigenvalues}
  p_j(t)=\frac{1+\nu_j(t)}{2}\quad{\rm and}\quad 1-p_j(t)=\frac{1-\nu_j(t)}{2}
\end{equation}
and entropy
$$
  S(\overline{\rho}_j)=H\left(\nu_j(t)\right)\,,
$$
where we define the function
\begin{equation} \label{h_fun:eqn}
  H(x)\equiv -\frac{1+x}{2}\log\left(\frac{1+x}{2}\right)-\frac{1-x}{2}\log\left(\frac{1-x}{2}\right)\,.
\end{equation}
Being the objects in Eq.~\ref{eigenvalues} the eigenvalues of a density
matrix, they are real and are in the interval $[0,1]$; this implies that $\nu_j\in[-1,1]$. 
The spectrum of $\rho_l$ results now from the $l$-fold product of the spectra of the density
matrices $\overline{\rho}_j$, and the entropy of $\rho_l$ is the sum of the entropies of the $l$ uncorrelated modes,
\begin{equation} \label{s_l:eqn}
  S_l(t)=\sum_{j=1}^l H\left(\nu_j(t)\right)\,.
\end{equation}
The evaluation of the correlation matrix and the entanglement entropy can be easily implemented numerically. 

%


\section{- Relaxation of the entanglement entropy to an asymptotic periodic regime.} \label{relaxation:sec}
Here we specialize the considerations of the section above to the case of periodic driving with period $\tau=2\pi/\omega_0$.
Expanding the Floquet states $\ket{\psi_k^{\pm}(t)}$ in the basis $\{ \opcdag{k} \opcdag{-k} \ket{0}, \ket{0} \}$ (see Sec.~\ref{Ising:sec})
we can find the Floquet solutions of the Bogoliubov-de Gennes equations Eq.~\eqref{deGennes:eqn}
\begin{equation} \label{psi_components:eqn}
  \psi_{F,\,k}^+(t)\equiv\left(\begin{array}{c}u_{P,\,k}(t)\\v_{P,\,k}(t)\end{array}\right)\nep^{-i\mu_k t},\quad{\rm and}\quad
    \psi_{F,\,k}^-(t)\equiv\left(\begin{array}{c}-v_{P,\,k}^*(t)\\u_{P,\,k}^*(t)\end{array}\right)\nep^{i\mu_k t}\,,
\end{equation}
where $u_{P,\,k}(t)$ and $v_{P,\,k}(t)$ are $\tau-$periodic objects.
Expanding Eq.~\eqref{expansion:eqn} in the basis $\left\{\opcdag{k}\opcdag{-k}\ket{0},\,\ket{0}\right\}$, we can write
a generic solution of the Bogoliubov-de Gennes equations as
\begin{equation}
  \left(\begin{array}{c}u_{k}(t)\\v_{k}(t)\end{array}\right) = 
     r_k^+\left(\begin{array}{c}u_{P,\,k}(t)\\v_{P,\,k}(t)\end{array}\right)\nep^{-i\mu_k t}+
      r_k^-\left(\begin{array}{c}-v_{P,\,k}^*(t)\\u_{P,\,k}^*(t)\end{array}\right)\nep^{i\mu_k t}\,.
\end{equation}
Substituting in Eq.~\eqref{R_cacca:eqn}, we find
\begin{eqnarray} \label{R_caccona:eqn}
  &&\hspace{-2cm}R_l(t)=\frac{1}{2\pi}\int_{-\pi}^\pi \nep^{-ikl} \left(R_{k}^{\,(\infty)}(t)+R_{k}^{\,({\rm fluc})}(t)\right)\ud k\,\quad
  I_l(t)=\frac{1}{2\pi}\int_{-\pi}^\pi \nep^{-ikl} \left(I_{k}^{\,(\infty)}(t)+I_{k}^{\,({\rm fluc})}(t)\right)\ud k\,,\nonumber\\
  &&\hspace{-2cm}Q_l(t)=\frac{1}{2\pi}\int_{-\pi}^\pi \nep^{-ikl} \left(Q_{k}^{\,(\infty)}(t)+Q_{k}^{\,({\rm fluc})}(t)\right)\ud k\,,
\end{eqnarray}
with
\begin{eqnarray} \label{infinitiva:eqn}
  &&\hspace{-2cm}R_{k}^{\,(\infty)}(t)\equiv\left(1-2|r_k^-|^2\right)\Real\left(u_{P,\,k}(t)v_{P,\,k}^*(t)\right),\quad
    R_{k}^{({\rm fluc})}(t)\equiv\Real\left(\left(u_{P,\,k}^2(t)-v_{P,\,k}^2(t)\right)r_k^+{r_k^-}^*\nep^{-2i\mu_k t}\right)\nonumber\\
  &&\hspace{-2cm}I_{k}^{\,(\infty)}(t)\equiv\left(1-2|r_k^-|^2\right)
              \Aimag\left(u_{P,\,k}(t)v_{P,\,k}^*(t)\right),\quad 
    I_{k}^{({\rm fluc})}(t)\equiv
               \Aimag\left(\left(u_{P,\,k}^2(t)-v_{P,\,k}^2(t)\right)
                  r_k^+{r_k^-}^*\nep^{-2i\mu_k t}\right)\nonumber\\
  &&\hspace{-2cm}Q_{k}^{\,(\infty)}(t)\equiv (2|r_k^+|^2-1)\left(|u_{P,\,k}(t)|^2-|v_{P,\,k}(t)|^2\right)
              \,,\quad 
    Q_{k}^{({\rm fluc})}(t)\equiv
               -4\Real\left(r_k^-{r_k^+}^*u_{P,\,k}^*(t)v_{P,\,k}^*(t)\nep^{-2i\mu_k t}\right)\,.\nonumber\\
\end{eqnarray}
When $t$ is large, the terms with superscript $({\rm fluc})$ are wildly oscillating in $k$ because of the factor $\nep^{-2i\mu_k t}$: integrating
them over $k$ we obtain a vanishingly small contribution. Only the terms with superscript $(\infty)$ survive: they are objects periodic with
period $\tau$. Therefore, asymptotically, the correlation matrix becomes the $\tau$-periodic object
\begin{equation} \label{corr_inf_mat:eqn}
 \Gamma_{\infty\,l}^C(t)=\left(\begin{array}{cccc}\Pi_{0}^{\,(\infty)}(t)&\Pi_{-1}^{\,(\infty)}(t)&\cdots&\Pi_{1-l}^{\,(\infty)}(t)\\
                                      \Pi_{1}^{\,(\infty)}(t)&\Pi_{0}^{\,(\infty)}(t)&&\vdots\\
                                      \vdots&&\ddots&\vdots\\
                                      \Pi_{l-1}^{\,(\infty)}(t)&\cdots&\cdots&\Pi_0^{\,(\infty)}(t)\end{array}\right)\,,
\end{equation}
\begin{equation} 
 \Pi_{l}^{\,(\infty)}(t)=\left(\begin{array}{cc}-2i{R}_{l}^{\,(\infty)}(t)  &  {Q}_{l}^{\,(\infty)}(t)-2iI_{l}^{\,(\infty)}(t)\\
                            -{Q}_{-l}^{\,(\infty)}(t)-2i I_{-l}^{\,(\infty)}(t)  &  2i{R}_{l}^{\,(\infty)}(t)\end{array}\right)\,,
\end{equation}
\begin{eqnarray}
 &&\hspace{-2cm}R_{l}^{\,(\infty)}(t)=\frac{1}{2\pi}\int_{-\pi}^\pi \nep^{-ikl} R_{k}^{\,(\infty)}(t)\,\ud k\,,\quad
 I_{l}^{\,(\infty)}(t)=\frac{1}{2\pi}\int_{-\pi}^\pi \nep^{-ikl} I_{k}^{\,(\infty)}(t)\,\ud k\,,\nonumber\\
 &&\hspace{-2cm}Q_{l}^{\,(\infty)}(t)=\frac{1}{2\pi}\int_{-\pi}^\pi \nep^{-ikl} Q_{k}^{\,(\infty)}(t)\,\ud k\,.
\end{eqnarray}
We see indeed that the correlation matrix relaxes asymptotically to a periodic value, and the same does the entanglement entropy. If the
eigenvalues of the matrix $i\Gamma_{\infty\,l}^C(t)$ are $\pm\nu_m^{(\infty)}(t)$, $m=1,\ldots,l$ (they are real and $\in[-1,1]$ as we show
in~\ref{entroval_subsec}), the entropy asymptotically becomes
\begin{equation} \label{superentropia:eqn}
  S_l^{\,(\infty)}(t) = \sum_{m=1}^l H\left(\nu_m^{(\infty)}(t)\right)\,,
\end{equation}
with $H(x)$ defined in Eq.~\eqref{h_fun:eqn}. 

\subsection{ - Numerical implementations} \label{num_imp:sec}
In the finite-time calculations of Fig.~\ref{entroppo:fig} 
we always implement the general form of the correlation matrix Eq.~\eqref{gener_corr:eqn}. 
Although the form given by Eqs.~\eqref{corr_mat_tras_t:eqn}
and~\eqref{R_caccona:eqn} may seem more appropriate for the PBC translationally invariant case, it is highly inconvenient. This is due
to the integrands $R_k(t)$, $I_k(t)$ and $Q_k(t)$, rapidly becoming too fast oscillating functions of $k$ because of the $\nep^{-i\mu_k t}$ factors.
To evaluate the correlation matrix Eq.~\eqref{gener_corr:eqn}, we have indeed to numerically solve the generic form of the 
Bogoliubov-de Gennes equations Eq.~\eqref{bog}: with our resources this is possible in a reasonable time for $L$ up to
$\sim800$. This is the reason why we cannot follow the convergence of the entanglement entropy in Fig.~\ref{entroppo:fig}
for very long times: after a time scaling linearly with $L$, finite size effects appear. The physical motivation of the finite size effects
lies in the propagation of quasi-particles at finite velocity, as detailed in~\cite{Russomanno_JSTAT13}.

On the opposite -- considering the the asymptotic entropy $S_l^{(\infty)}(0)$ (Eq.~\eqref{superentropia:eqn}) --
$R_k^{(\infty)}(0)$, $I_k^{(\infty)}(0)$ and $Q_k^{(\infty)}(0)$ (see Eq.~\eqref{infinitiva:eqn}) are smooth functions of $k$ whose integral be easily
evaluated even with moderate values of $L$~\footnote{We perform the integrals in $k$ by means of a cubic interpolation~\cite{Recipes:book} and the use of the quadpack integration routines~\cite{quadpack:book}.} (and then not too tight meshes in $k$ -- there is some {\em caveat} for small frequencies
which we discuss in Sec.~\ref{numerically:sec}). Indeed, Eq.~\eqref{corr_inf_mat:eqn} is very suitable for
the numerical evaluation of the asymptotic entropy. The resulting asymptotic entropy is appropriate
both for the OBC case and the PBC one: the reason is the argument based on the propagation of quasi-particles that 
we explain in Sec.~\ref{numerically:sec}. 

For large enough $l$, the asymptotic entanglement 
entropy scales linearly in $l$ (see Eq.~\eqref{sel_leading:eqn}) as we are going to show.

%
\section{- Semi-analytical approximation of $S_l^{\,(\infty)}(t)$, for large $l$} \label{Appendix-semi}
In this appendix, we are going to explicitly demonstrate Eq.~\eqref{sel_leading:eqn}. 

%
%
We perform an analysis very similar to the one introduced by Calabrese and Cardy in Ref.~\cite{Calabrese_JSTAT05}~\footnote{It is
not difficult to see that our Eq.~\eqref{sel_leading:eqn} reduces to Eq. (3.19) of Ref.~\cite{Calabrese_JSTAT05} if we consider
a quantum quench as a degenerate periodic driving of arbitrary period.}.
We define
\begin{eqnarray}
   \Pi_{k}^{\,(\infty)}(t)&=&\left(\begin{array}{cc}-2i{R}_{k}^{\,(\infty)}(t)&{Q}_{k}^{\,(\infty)}(t)-2iI_{k}^{\,(\infty)}(t)\\
                            -{Q}_{k}^{\,(\infty)}(t)-2i I_{k}^{\,(\infty)}(t)&2i{R}_{k}^{\,(\infty)}(t)\end{array}\right)\,,\nonumber\\
   \widetilde{{\Gamma}}^{C}_{\infty\,l}(t,\lambda) &=& i\lambda\boldsymbol{1}_{2l} - {\Gamma}^{C}_{\infty\,l}(t)\quad{\rm and}\nonumber\\
   e_2(x,y) &=& -\frac{y+x}{2}\log\left(\frac{y+x}{2}\right) - \frac{y-x}{2}\log\left(\frac{y-x}{2}\right)\,.
\end{eqnarray}
Being $\pm\nu_m$~\footnote{In this section we omit the superscript ``$(\infty)$'' and we do not make explicit the dependence
on $t$ in the $\pm\nu_m$ in order to have a simpler notation.} 
the eigenvalues of $i{\Gamma}^{C}_{\infty\,l}(t)$, 
we see that the eigenvalues of the matrix $\widetilde{{\Gamma}}^{C}_{\infty\,l}(t,\lambda)$ are $i\lambda\pm i\nu_j$.
Its determinant is therefore
\begin{displaymath}
  D_{\,l\,t}(\lambda) \equiv \det \widetilde{{\Gamma}}^{C}_{\infty\,l}(t,\lambda) = (-1)^{l}\prod_{j=1}^l\left(\lambda^2-\nu_j^2\right)\,.
\end{displaymath}
Applying the Cauchy theorem to Eq.~\eqref{superentropia:eqn}, we find that we can write the asymptotic entanglement 
entropy $S_l^{\,(\infty)}(t)$ as~\cite{Calabrese_JSTAT05,Jin_JStP04,Jin_JPA05}
\begin{equation} \label{sel_mezzo:eqn}
  S_l^{\,(\infty)}(t) = \frac{1}{4\pi i}\oint_c\left(\frac{1}{\lambda-\nu_j}+\frac{1}{\lambda+\nu_j}\right)e_2(1+\epsilon,\lambda)\ud\lambda=
              \frac{1}{4\pi i}\oint\ud\lambda\, e_2(1+\epsilon,\lambda)\frac{\ud}{\ud\lambda}\log D_{\,l\,t}(\lambda)
\end{equation}
(the second equality can be verified by direct substitution). The contour encircles the interval $[-1,1]$ where there are the poles of 
 $\frac{\ud}{\ud\lambda}\log D_{\,l\,t}(\lambda)$; it tends to this interval as $\epsilon\to 0$. The $\epsilon$ in the first argument
of $e_2$ is necessary to make the branch cuts of $e_2$ to be outside the integration contour: inside the contour the function
is analytical and this justifies the application of the Cauchy theorem.

\textcolor{black}{Refs.~\cite{Calabrese_JSTAT05, MCCoy_PRB74} show that, when $l\gg 1$, being $\widetilde{{\Gamma}}^{C}_{\infty\,l}(t,\lambda)$ a Toeplitz matrix, the
approximate formula
\begin{equation}
  \log\det \widetilde{{\Gamma}}^{C}_{\infty\,l}(t,\lambda) = 
     \frac{l}{2\pi}\int_{-\pi}^\pi\log\det\left(i\boldsymbol{1}_2\lambda-{\Pi}_k^{\,(\infty)}(t)\right)\ud k+\mathcal{O}(\log l)
\end{equation}
is valid. 
Notice that $t$, in the previous formula, does not scale with $l$: since 
$\widetilde{{\Gamma}}^{C}_{\infty\,l}(t,\lambda)$ is a $\tau$-periodic quantity we can restrict $t$ to the interval 
$[0,\tau]$.
The situation is very similar to Ref.~\cite{Calabrese_JSTAT05} where the authors apply the same formula when they consider an asymptotic  static condition after a quantum quench.} 
%
%
Using the definitions we find
\begin{equation} \label{deterambo:eqn}
  \det\left(i\lambda\boldsymbol{1}_2-{\Pi}_k^{\,(\infty)}(t)\right) = 
     -\lambda^2+4(R_k^{\,(\infty)}(t))^2+4(I_k^{\,(\infty)}(t))^2+(Q_k^{\,(\infty)}(t))^2 = -\lambda^2+A_k^2\,,
\end{equation}
where we have defined $A_k\equiv(1-2|r_k^-|^2)$ which is remarkably {\em time-independent}. 
We see that $4(R_k^{\,(\infty)}(t))^2+4(I_k^{\,(\infty)}(t))^2+(Q_k^{\,(\infty)}(t))^2=A_k^2<1$ because this is the norm of the Bloch-vector representative of the Floquet GGE density matrix (Eq.~\eqref{periodic_GGE:eqn}) which can also be written as
\begin{displaymath}
\rho_k^{\,FGGE}(t)\equiv\left(\begin{array}{cc}Q_k^{\,(\infty)}(t)/2&R_k^{\,(\infty)}(t)+iI_k^{\,(\infty)}(t)\\
                                                    R_k^{\,(\infty)}(t)-iI_k^{\,(\infty)}(t)&-Q_k^{\,(\infty)}(t)/2\end{array}\right)\,.
\end{displaymath}
%
%
%
We find indeed (omitting the logarithmic corrections)
\begin{equation}
  \frac{\ud}{\ud\lambda}\log D_{\,l}(\lambda) = 
     \frac{l}{2\pi}\int_{-\pi}^\pi\ud k \frac{\ud}{\ud\lambda}\log\left(-\lambda^2+A_k^2\right) =
       \frac{l}{2\pi}\int_{-\pi}^\pi\ud k \frac{2\lambda}{\lambda^2 - A_k^2}\,.
\end{equation}
Substituting in Eq.~\eqref{sel_mezzo:eqn} and using the value of $A_k$, we find
\begin{equation} \label{sel_pazzo:eqn}
  S_l =  \frac{l}{2\pi}\int_{-\pi}^\pi\ud k\frac{1}{4\pi i}\oint\ud\lambda\, e_2(1+\epsilon,\lambda)\frac{2\lambda}{\lambda^2 - A_k^2}\,.
\end{equation}
%
Being $A_k^2<1$, both the poles fall inside
the integration contour and we can explicitly perform the integral in $\ud\lambda$ Eq.~\eqref{sel_pazzo:eqn}. After this straightforward
integration, we send $\epsilon\to 0$ and find Eq.~\eqref{sel_leading:eqn}
\begin{eqnarray}
  &&\hspace{-2.3cm}S_l^{\,(\infty)}(t) = l\, s^{({\infty})}+\mathcal{O}(\log l)\quad{\rm with}\quad\nonumber\\
     &&\hspace{-2cm}s^{({\infty})} \equiv \frac{1}{2\pi}\int_{-\pi}^\pi\ud k\, H\left(A_k\right)=
                   -\frac{1}{\pi}\int_{0}^\pi\ud k\, \left[|r_k^-|^2\log|r_k^-|^2+|r_k^+|^2\log|r_k^+|^2\right] \,,\nonumber
\end{eqnarray}
where $H(x)$ is defined in Eq.~\eqref{h_fun:eqn} and we have used the symmetry $|r_k^-|^2=|r_{-k}^-|^2$ which is implied by the Hamiltonian matrix in Eq.~\eqref{Ht:eqn} obeying the 
relation $\mathbb{H}_{-k}(t)=\sigma_z\mathbb{H}_{k}(t)\sigma_z$~\cite{russomanno_JSTAT15}.

We stress that we cannot apply the foregoing argument for $t$ finite: in that case we have $A_k=1$ for all $k$. All the poles in 
Eq.~\eqref{sel_pazzo:eqn} coincide and 
this analysis is not suitable. 
That is so because in this case $A_k^2=4(R_k(t))^2+4(I_k(t))^2+(Q_k(t))^2=1$ is
the norm of the Bloch vector of the pure state $\left(\begin{array}{c}u_k(t)\\v_k(t)\end{array}\right)$. 
\section*{Bibliography}
\vspace{3mm}
%

\end{document}